 \documentclass[pmlr,twocolumn,10pt]{jmlr} % W&CP article

 \usepackage[mode=buildnew]{standalone}

\usepackage{booktabs}
 
\usepackage[load-configurations=version-1]{siunitx} % newer version 
\usepackage{amsmath}

\usepackage{amsfonts}

\theorembodyfont{\upshape}
\theoremheaderfont{\scshape}
\theorempostheader{:}
\theoremsep{\newline}

\jmlrvolume{LEAVE UNSET}
\jmlryear{2021}
\jmlrsubmitted{LEAVE UNSET}
\jmlrpublished{LEAVE UNSET}
\jmlrworkshop{Machine Learning for Health (ML4H) 2021} % W&CP title

\title[Interpretability Aware Model Training]{Interpretability Aware Model Training to Improve Robustness against Out-of-Distribution Magnetic Resonance Images in Alzheimer's Disease Classification}

\author{
  \Name{Merel Kuijs}
  \Email{kuijsm@student.ethz.ch}\\ 
  \addr Machine Learning \& Computational Biology, Department of Biosystems Science and Engineering,\\
  ETH Zurich, Basel, Switzerland\\
  Institute of Computational Biology, Helmholtz Zentrum München, Neuherberg, Germany 
  \AND
  \Name{Catherine R. Jutzeler}
  \Email{catherine.jutzeler@bsse.ethz.ch}\\ 
  \addr Machine Learning \& Computational Biology, Department of Biosystems Science and Engineering,\\
  ETH Zurich, Basel, Switzerland\\
  Swiss Institute for Bioinformatics (SIB), Lausanne, Switzerland
  \AND
  \Name{Bastian Rieck}
  \Email{bastian.rieck@bsse.ethz.ch}\\ 
  \addr Machine Learning \& Computational Biology, Department of Biosystems Science and Engineering,\\
  ETH Zurich, Basel, Switzerland\\
  Swiss Institute for Bioinformatics (SIB), Lausanne, Switzerland\\
  Institute of AI for Health, Helmholtz Zentrum München, Neuherberg, Germany 
  \AND
  \Name{Sarah C. Br\"uningk}
  \Email{sarah.brueningk@bsse.ethz.ch}\\ 
  \addr Machine Learning \& Computational Biology, Department of Biosystems Science and Engineering,\\
  ETH Zurich, Basel, Switzerland\\
  Swiss Institute for Bioinformatics (SIB), Lausanne, Switzerland
}
\begin{document}

\maketitle

\begin{abstract}
Owing to its pristine soft-tissue contrast and high resolution, structural magnetic resonance imaging (MRI) is widely applied in neurology, making it a valuable data source for image-based machine learning (ML) and deep learning applications. The physical nature of MRI acquisition and reconstruction, however, causes variations in image intensity, resolution, and signal-to-noise ratio. Since ML models are sensitive to such variations, performance on out-of-distribution data, which is inherent to the setting of a deployed healthcare ML application, typically drops below acceptable levels. We propose an interpretability aware adversarial training regime to improve robustness against out-of-distribution samples originating from different MRI hardware. The approach is applied to 1.5T and 3T MRIs obtained from the Alzheimer’s Disease Neuroimaging Initiative database. We present preliminary results showing promising performance on out-of-distribution samples. 
\end{abstract}

\section{Introduction}
\label{sec:intro}

Magnetic resonance (MR) imaging produces high resolution, high contrast, three dimensional (3D) anatomical representations based on local variations in magnetic susceptibility \citep{duyn2007high}. Scan quality, quantified by the signal-to-noise ratio and image resolution, varies depending on the imaging sequence, hardware (e.g.\ magnetic field strength, choice of coils, and shimming correction), and reconstruction software \citep{rutt1996impact,LADD20181} in addition to tissue characteristics \citep{bloem2018mr}. In clinical practice, 1.5T scanners provide a reasonable compromise between scan quality and imaging/machine cost. Improved signal-to-noise ratios are achieved with 3T scanners, which are therefore preferable if a high level of morphological detail is required, e.g.\ in neurological studies of the brain. Differences between scans originating from different hard-/software are generally subtle and usually would not impact diagnosis by a human observer. However, deep learning models are very vulnerable to small shifts in the data distribution and non-robust models do not generalize to such out-of-distribution (OOD) samples. Models trained on images acquired using the same protocol and scanner generalize poorly to data acquired using different imaging protocols or hardware \citep{Allen2019,Bluemke2020,lee2020generalization}. Models perform particularly poorly on scans of lower quality than that used for training \citep{guan2021multi}. In practice, real world data is acquired using a variety of scanners and protocols in different clinics. The training data will therefore never be representative of all examples a model may encounter upon deployment. So far, only models trained with multiple curated data sets or strongly augmented data have been able to achieve some amount of generalization \citep{Chen2020,maartensson2020reliability,Zhang2020}. A similar challenge is provided by $L_\infty$ adversarial attacks, which disturb input data minimally to produce substantial changes in the output of the classifier. Since individual pixel values only change subtly, the perturbations are often invisible to the human observer, but will fool a classifier nevertheless. Recently, \citet{boopathy2020proper} have shown that adversarial examples which successfully fool a classifier often fail to prevent 2-class interpretation discrepancies from occurring. Penalizing a network based on 2-class interpretation discrepancies during training improves adversarial robustness. We hypothesize that robustness to adversarial attacks could also increase OOD robustness, hence improving the generalization ability of medical image classifiers. This idea motivated our study, which investigates interpretability aware adversarial training as a means to improve OOD robustness in the realm of MR imaging. This preliminary analysis applies the presented concept to Alzheimer's disease classification. We train convolutional neural networks (CNNs) on high quality MR images acquired at a specific magnetic field strength (3T) and test on lower quality OOD images (1.5T). 

\section{Methods}
\label{sec:methodology}
\subsection{Model}
Following \citet{bruningk2021back}, we use simple 2D and 3D CNNs comprising three convolutional layers with batch normalization, ReLU activation and max pooling, followed by a fully connected layer. Dropout is applied to the flattened outputs of the last convolutional layer. L2 regularization was used on the weights of all layers. Hyperparameter optimization is described in \ref{HP_optimization}.

\setlength{\belowdisplayskip}{0pt} 
\setlength{\belowdisplayshortskip}{0pt}
\setlength{\abovedisplayskip}{0pt} 
\setlength{\abovedisplayshortskip}{0pt}

\subsection{Interpretability Measure}
Grad-CAM is a widely-used posthoc interpretability method to visualize the evidence on which a classifier bases its decisions \citep{selvaraju2017grad}. In contrast to other approaches such as CAM \citep{zhou2016learning}, it does not constrain the model architecture. Given a class-specific prediction $y^c$, Grad-CAM calculates weights $n_m^c$, $m \in \{1, \dots, n\}$, to weigh the (3D) convolutional output maps $f_1, \dots, f_n$,
\begin{equation}
    n_m^c = \frac{1}{|xyz|} \sum_x\sum_y\sum_z \frac{\partial y^c}{\partial f_{xyz}^m}.
\end{equation}
The output maps are multiplied by the corresponding weights to arrive at the class activation map,
\begin{equation}
    I^c = ReLu\left(\sum_m n_m^cf^m\right). \label{eq:map}
\end{equation}

\subsection{Training Regimes}
We compared three baseline training regimes (normal, combined, and adversarial) to the proposed interpretability aware adversarial approach. The normal and combined baselines minimize the \textit{standard} binary cross-entropy loss function $L$,
\begin{align}
    L = \frac{-1}{n}\sum_{i}^{n} y_i \cdot \log \hat{y}_i + (1 - y_i) \cdot \log(1 - \hat{y}_i)
    \label{eq:standard_CEloss}
\end{align}
where $y_1 \dots y_n$ are the labels associated with the benign examples $x_1 \dots x_n$ and $\hat{y}_i \dots \hat{y}_n$ are the corresponding predictions.\\
Models trained under the adversarial regime initially use the \textit{standard} cross-entropy loss function (\ref{eq:standard_CEloss}). After 400 steps, we include adversarial examples generated using the projected gradient descent (PGD) algorithm \citep{madry2017towards} and apply an \textit{adversarial} cross-entropy loss function $L_{adv}$,
\begin{align}
\begin{split}
    L_{adv} =& \frac{-1}{n}\sum_{i}^{n} y_i \cdot \log \hat{y}_{adv,i} +\\
    &(1 - y_i) \cdot \log(1 - \hat{y}_{adv,i}),
    \label{eq:adv_CEloss}
\end{split}   
\end{align}
where $\hat{y}_{adv,1} \dots \hat{y}_{adv,n}$ are the adversarial predictions. Very briefly, the PGD attack adds disturbances that depend on an iterative gradient calculation. The disturbances are clipped to the $[-\epsilon, \epsilon]$ domain to control the strength of the attack. For training, we first linearly increase $\epsilon$ over 2000 steps. Training then continues at maximum attack strength until early stopping (scoring validation loss) is triggered. We used a custom learning rate scheduler, too (20\% learning rate reduction upon plateau). We evaluated perturbation sizes $\epsilon \in \{0.001, 0.005\}$.\\
We define a measure of the $\ell_1$ 2-class interpretation discrepancy
\begin{align}
\begin{split}
    \tilde{D}_{2, \ell_1}\left(x, x_{adv}\right) =& \frac{1}{2}\left( \|I(x,c_1) - I(x_{adv},c_1)\|_1 + \right. \\
    & \left. \|I(x,c_2) - I(x_{adv},c_2)\|_1 \right),
\label{eq:int_disc}
\end{split}
\end{align}
where $I(x,c)$ denotes a class-specific (either class $c_1$ or $c_2$) activation map generated by Grad-CAM following \equationref{eq:map}. 
% We define a measure of the worst-case interpretation discrepancy $\tilde{D}_{worst}$ as follows:
% \begin{align}
%     &\tilde{D}_{worst}\left(x, x_{adv}\right) =  
%     &\tilde{D}_{2, \ell_1}\left(x, x + \argmax_{\|\delta\|_\infty \leq \epsilon} [L\left(\theta, x + \delta, y\right)] \right),
% \end{align}
% where $\theta$ represents a model fit parameter. $L\left(\theta, x + \delta, y\right)$ is the training loss (i.e. the \textit{adversarial} cross-entropy loss defined in \ref{adv_CEloss}). 
Training of the interpretability aware model follows the process outlined above for adversarial training but now the model switches to an \textit{adjusted} loss $L_{adj}$ combining the \textit{adversarial} cross-entropy loss with the measure of the $\ell_1$ 2-class interpretation discrepancy, 
\begin{align}
    L_{adj} = L_{adv} + \lambda \tilde{D}_{2, \ell_1}\left(x, x_{adv,\epsilon} \right),
    \label{eq:adj_CEloss}
\end{align}
where the regularization parameter $\lambda > 0$ controls the balance between performance on adversarial examples and class activation map similarity. We investigated values of $\lambda$ ranging from 1 to 30.

\begin{figure*}[htbp]
\floatconts
  {fig:metrics_1_out}
  {%
    \addtocounter{figure}{-1}
    \subfigure[TPR]{\label{fig:TPR_1_out}%
      \includegraphics[width=0.3\linewidth]{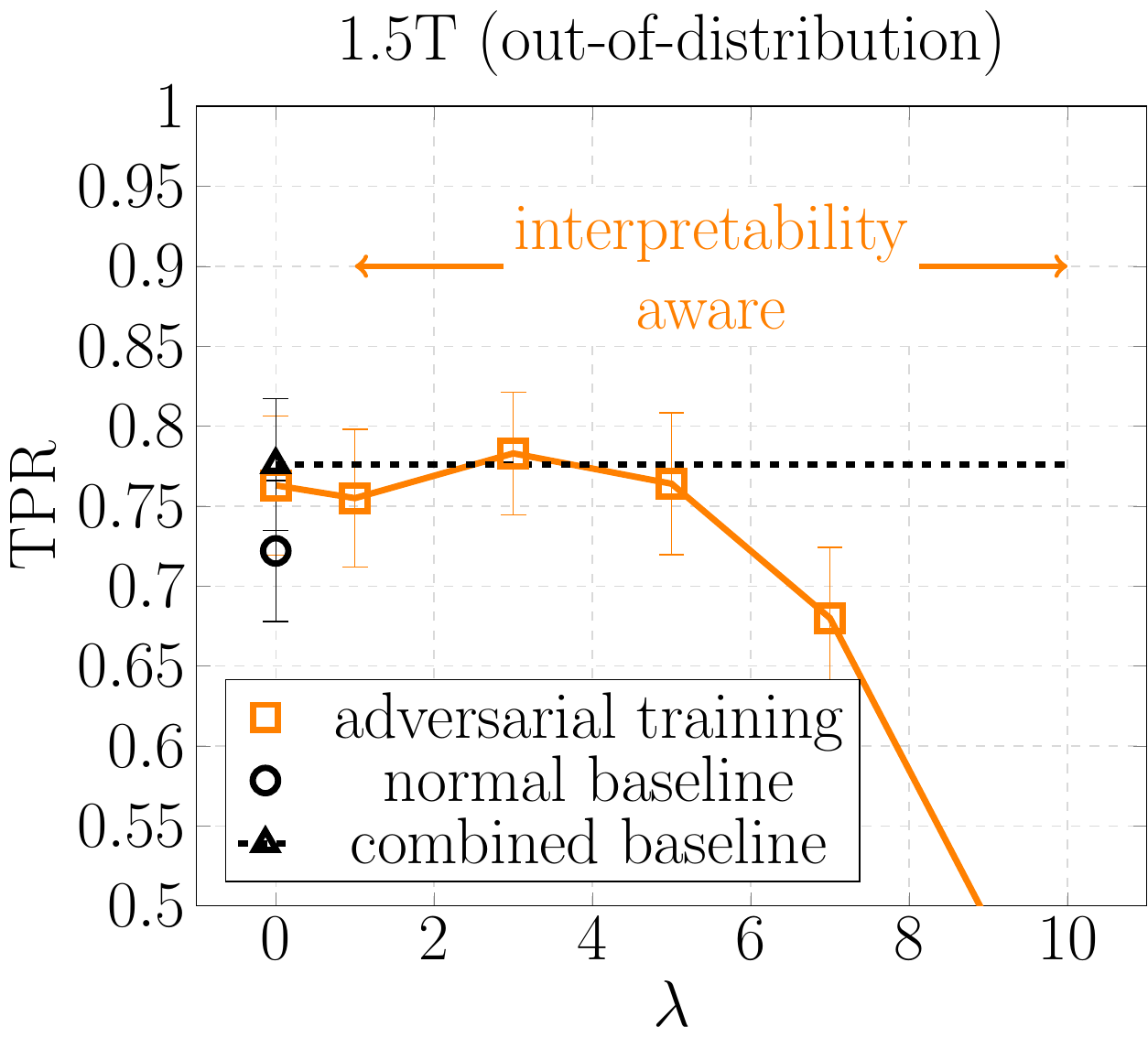}}%
    \qquad
    \subfigure[TNR]{\label{fig:TNR_1_out}%
      \includegraphics[width=0.3\linewidth]{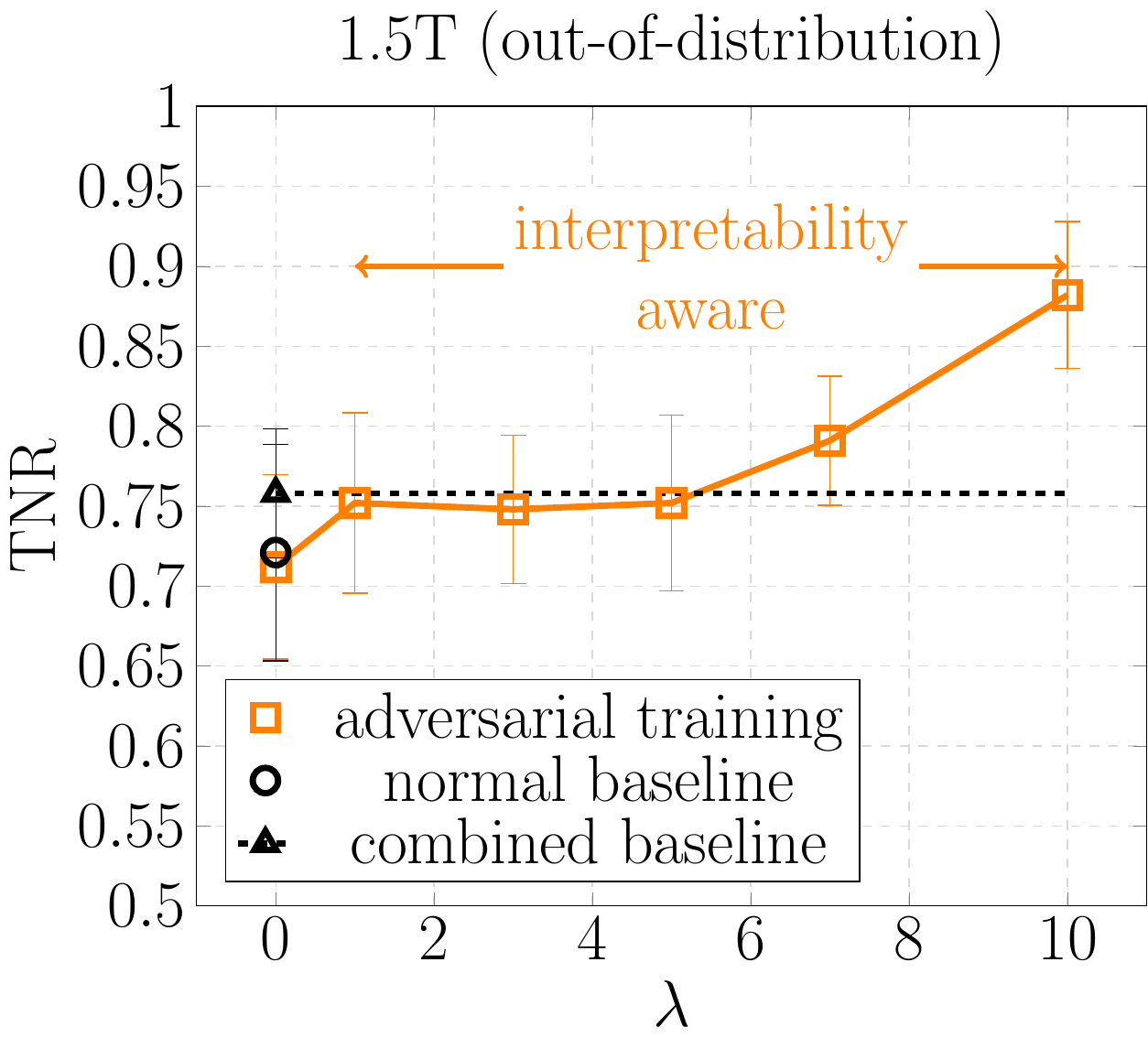}}
    \qquad
    \subfigure[ACC]{\label{fig:ACC_1_out}%
      \includegraphics[width=0.3\linewidth]{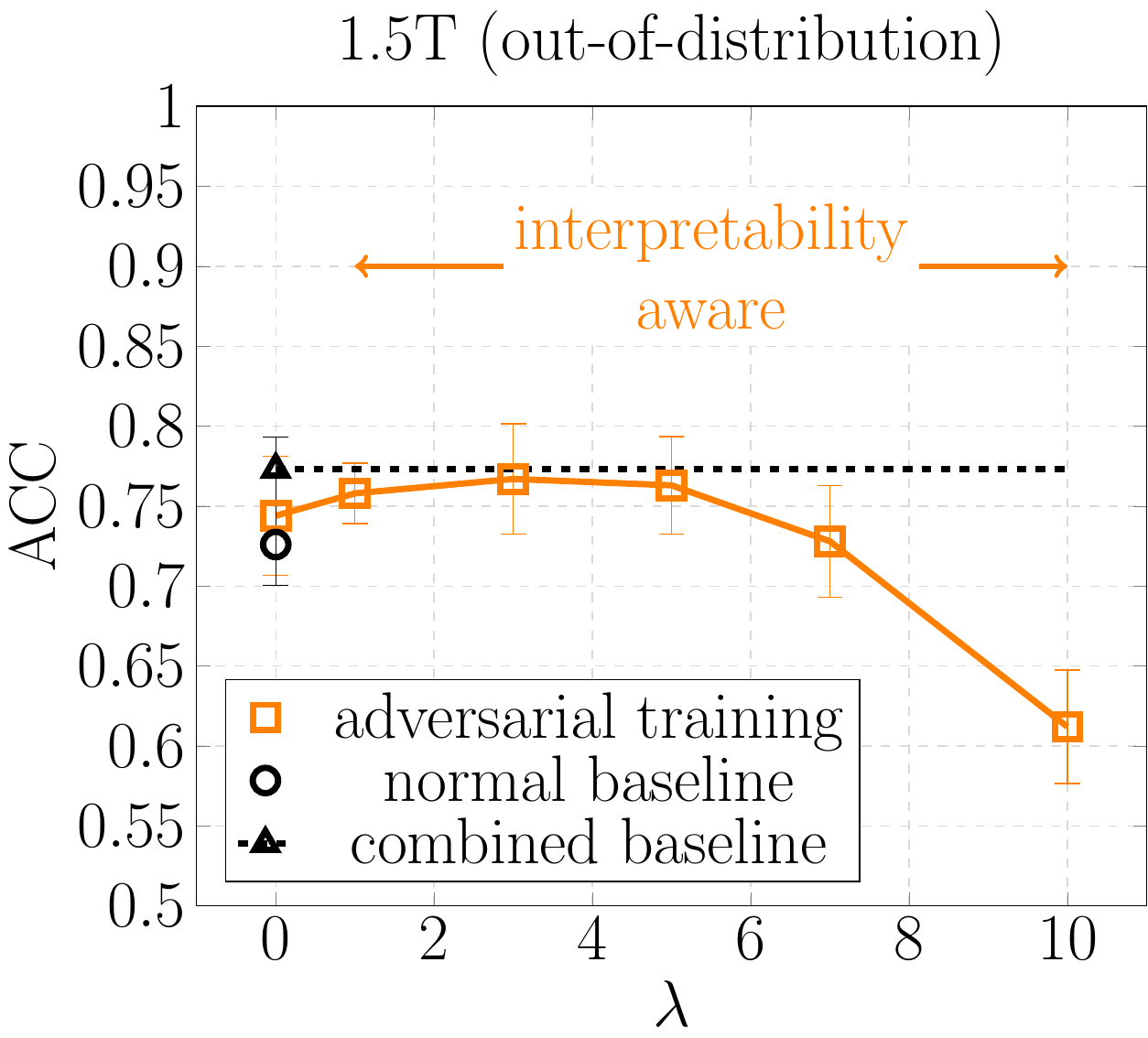}}
  }
  {\caption{\textbf{2D OOD} performance metrics shown as a function of $\lambda$ (no further improvement for $\lambda>5$). Performance is evaluated on undisturbed OOD (1.5T) data. The adversarial model ($\lambda = 0$) and the interpretability aware models ($\lambda > 0$) were trained using $\pmb{\epsilon = 0.001}$.}}
\end{figure*}

\section{First Experiments}
\subsection{Data}
\label{sec:data}
We test our approach on structural, T1-weighted MR images of Alzheimer's disease (AD) and cognitive normal (CN) subjects from the \href{http://adni.loni.usc.edu/}{Alzheimer's Disease Neuroimaging Initiative}\footnote{\texttt{http://adni.loni.usc.edu/}} (ADNI) database. 
% The ADNI was launched in 2003 as a public-private partnership, led by Principal Investigator Michael W. Weiner, MD. The primary goal of ADNI has been to test whether serial magnetic resonance imaging (MRI), positron emission tomography (PET), other biological markers, and clinical and neuropsychological assessment can be combined to measure the progression of mild cognitive impairment (MCI) and early Alzheimer’s disease (AD). 
1.5T (n=358, mean prevalence of AD=0.55) and 3T (n=247, mean prevalence of AD=0.35) images of non-overlapping subjects were included. Data pre-processing and image subset selection is described in \ref{data_prep}.

\subsection{Performance Evaluation}
The \textit{normal baseline}, the adversarial model, and the interpretability aware models were trained on 3T data (within-distribution, WD) whereas the \textit{combined baseline} trained on both 1.5T and 3T data. All models were evaluated on (i) undisturbed 3T data, (ii) adversarial examples generated from 3T data, and (iii) 1.5T data (OOD). We performed 5-fold nested cross validation and report results calculated on the held out test sets in Fig.~1. All results were averaged over all folds and three repeated runs to report means $\pm$ standard deviations. We evaluated test set prediction accuracy, TPR (sensitivity), TNR (specificity), and areas under the receiver operator (AUC) and precision recall curves (APS, i.e. AUPRC).

\section{Results}
\label{sec:results}
% adjust ticks and color scheme 

We first evaluate \textit{normal} and \textit{combined baseline} performance on OOD vs WD data in \tableref{tab:normal}. As expected, performance degrades if training and testing magnetic field strengths differ, but improves if both 1.5T and 3T scans are seen during training. \figureref{fig:TPR_1_out,fig:TNR_1_out,fig:ACC_1_out} compare the OOD performance of the 2D adversarial ($\lambda = 0$, $\epsilon_{train} = 0.001$) and interpretability aware models ($\lambda > 0$, $\epsilon_{train} = 0.001$) to that of the baselines. AUROC and AUPRC scores as well as WD metrics are provided in the appendix. In line with results reported by \citet{boopathy2020proper}, the interpretability aware model is more robust to adversarial examples compared to the model trained with adversarial attacks but without interpretation discrepancy penalties (\figureref{fig:TPR_1_in}--\figureref{fig:APS_1_in}). Increased adversarial robustness translated into a small improvement in OOD performance: 0.78 (interpretability aware, $\lambda = 3$) vs 0.76 (adversarial), 0.75 vs 0.71, and 0.77 vs 0.74 TPR, TNR, and ACC. The OOD performance of the best interpretability aware models approaches that of the \textit{combined baseline} trained on both 1.5T and 3T images. Compared to the \textit{normal baseline}, the best interpretability aware models ($\lambda \in \{3, 5\}$) perform significantly better (p\textless0.05, Mann-Whitney U test). These trends were observed both in the $\epsilon_{train} = 0.001$ and $\epsilon_{train} = 0.005$ (\figureref{fig:TPR_5_out}--\figureref{fig:APS_5_out}) setting. Preliminary results in the 3D setting (\figureref{fig:3D_TPR_1_out}--\figureref{fig:3D_APS_1_out}) show that the 3D interpretability aware model is superior to the adversarial model and the \textit{normal baseline} on the 1.5T data, too. Representative examples of Grad-CAM saliency maps of OOD examples obtained with and without interpretability aware training are shown in the appendix (\figureref{fig:tn_c0_int}--\ref{fig:tp_c1_adv}). In the latter case, we observe that clinically meaningful explanations highlighting tissue atrophy as AD evidence are obtained.

% numbers 
% Interestingly, the normal baseline trained on 3T data already generalizes relatively well to 1.5T data, and vice versa (\tableref{tab:normal}). Although the adversarial and the interpretability aware model have slightly better performance on AD samples -- the minority class in the train set -- as is indicated by better TPR scores, overall performance is not vastly different. \\

\section{Discussion}
\label{sec:discussion}
In the healthcare domain, model generalization to OOD data is an immanent challenge given the variety in hardware and post-processing algorithms. Additionally, model interpretability is essential to build trust in clinical ML predictions \citep{ahmad2018interpretable,carvalho2019machine} -- it is a natural next step to harness this concept for an alternative, intuitive method to increase model robustness. Following previous work \citep{boopathy2020proper}, we extended the proposed approach to be compliant with 2D/3D MR data, implemented Grad-CAM as interpretability measure, and, as a novel use case, demonstrate that this approach improves OOD robustness. Following careful hyperparameter optimization, we show that increased adversarial robustness translates into improved OOD robustness against images obtained using varying hardware. Interpretability aware training could hence be a complementary concept to (extreme) data augmentation and training on heterogeneous inputs, which were previously suggested to improve model generalizability \citep{Chen2020,maartensson2020reliability,Zhang2020}. Given its intuitive conceptualization, interpretability aware training may be well-accepted by the clinical community, too. Notably, the used data set was greatly harmonized, which may limit the potential benefit of interpretability aware adversarial training. Specifically, ADNI MR images were acquired using a unified protocol \citep{jack2008alzheimer} including scanner-specific pre-scan procedures and a phantom calibration scan prior to every patient imaging. The procedure was designed to harmonize scans and hence strongly exceeds the normal standards of clinical practice. Thus, the evaluation on the ADNI data should only be considered a preliminary analysis and we plan to test our approach in a more realistic setting using images from the UK Biobank\footnote{\texttt{www.ukbiobank.ac.uk}}, OASIS\footnote{\texttt{www.oasis-brains.org}} and the Health-RI Parelsnoer Neurodegenerative Diseases Biobank\footnote{\texttt{www.health-ri.nl/parelsnoer}} as true external validation sets. Moreover, we are planning to further investigate the use of different interpretability methods. Grad-CAM relies on up-sampling the final class activation map and hence suffers from poor resolution. Recently, LayerCAM has been proposed as an effective means to mitigate this limitation by weighting each image pixel using the backward class-specific gradients, yielding more fine-grained visualizations \citep{jiang2021layerCAM}. However, saliency map reproducibility and localization ability has previously been criticized in the context of complex models trained on medical image classification tasks \citep{arun2021assessing}. Here, we use a simple model driven by the clinical hallmarks of AD \citep{bruningk2021back}. Also, the model does \emph{not} interpret the meaningfulness of the Grad-CAM representations but uses a quantitative measure of salience map differences as a regularizing term only. % We further show that improved OOD robustness translated to more meaningful explanations, too (see appendix). 
Finally, a strong assumption of this study was that adversarial examples can be used to mimic the variation of MR images induced by physical processes. We provide preliminary data based on PGD attacks, but it is also important to test on unseen attacks and compare attacks in general. Examples could include Gabor, Snow \citep{kang2019testing}, and $\ell_1$ attacks \citep{chen2018ead}.\\
We have presented here a preliminary analysis of interpretability aware training to boost OOD performance on medical MR images which poses an important challenge in the field of ML for healthcare.

% \subsection*{Summary of the preliminary findings}
% - model performs ok (2D, 3D) on benign examples. 
% - emphasize the novelty of your approach (MRI data, Grad-CAM) but also honestly state that this approach was previously suggested. 

% \subsection*{Limitations of the current approach}
% To account for differences in hardware, protocols, and other factors that affect signal-to-noise ratio and image resolution, the ADNI consortium has developed harmonized protocols to minimize differences between images of different patients (and between images of the same patient taken at different times). This protocol includes specific requirements regarding scanner-specific imaging sequences and corresponding parameters. As such, the ADNI database provides the basis for automatic image classification tasks investigated using a variety of machine and deep learning approaches.

% - not very strong differences (ADNI data) 
% - only 3T-1.5T works
% - careful HP adaptation required
% - attack not mimicking the physical differences in MRIs
% - slow training

% \subsection*{Outlook and future direction}
% - layerCAM
% - different attacks
% - different data set 
\newpage
\acks
This work was partially funded and supported by the Swiss National Science Foundation (Spark grants 190466~(B.R.) and 190647~(S.C.B., C.R.J); Ambizione Grant \#PZ00P3186101, C.R.J). Since October 2021 the work of SCB was supported by the Botnar Research Centre for Child Health Postdoctoral Excellence Programme under PEP-2021-1008.
Data collection and sharing for this project was funded by the Alzheimer's Disease Neuroimaging Initiative (ADNI) (National Institutes of Health Grant U01 AG024904) and DOD ADNI (Department of Defense award number W81XWH-12-2-0012). ADNI is funded by the National Institute on Aging, the National Institute of Biomedical Imaging and Bioengineering, and through generous contributions from the following: AbbVie, Alzheimer's Association; Alzheimer's Drug Discovery Foundation; Araclon Biotech; BioClinica, Inc.; Biogen; Bristol-Myers Squibb Company; CereSpir, Inc.; Cogstate; Eisai Inc.; Elan Pharmaceuticals, Inc.; Eli Lilly and Company; EuroImmun; F. Hoffmann-La Roche Ltd and its affiliated company Genentech, Inc.; Fujirebio; GE Healthcare; IXICO Ltd.; Janssen Alzheimer Immunotherapy Research \& Development, LLC.; Johnson \& Johnson Pharmaceutical Research \& Development LLC.; Lumosity; Lundbeck; Merck \& Co., Inc.; Meso Scale Diagnostics, LLC.; NeuroRx Research; Neurotrack Technologies; Novartis Pharmaceuticals Corporation; Pfizer Inc.; Piramal Imaging; Servier; Takeda Pharmaceutical Company; and Transition Therapeutics. The Canadian Institutes of Health Research is providing funds to support ADNI clinical sites in Canada. Private sector contributions are facilitated by the Foundation for the National Institutes of Health (www.fnih.org). The grantee organization is the Northern California Institute for Research and Education, and the study is coordinated by the Alzheimer's Therapeutic Research Institute at the University of Southern California. ADNI data are disseminated by the Laboratory for Neuro Imaging at the University of Southern California.

\newpage
\bibliography{jmlr-sample.bib}

\newpage
\appendix
\renewcommand\thefigure{S\arabic{figure}} 
\renewcommand\thetable{S\arabic{table}} 

\section{Supplementary Methods}
\subsection{Data preparation}
\label{data_prep}
All images were preprocessed using the \href{https://fmriprep.org/en/stable/}{\texttt{fmriprep}}\footnote{\texttt{https://fmriprep.org/en/stable/}} pipeline for bias-field correction, reference space registration, and skull stripping, as previously described in \citet{bruningk2021back}. Images were intensity normalized (mean zero and unit variance) and segmented into functional brain subunits using the CerebrA atlas \citep{manera2020cerebra}. It was recently demonstrated that constraining the model input to clinically relevant brain subunits boosted performance \citep{bruningk2021back}. Hence, we train our models on the image subset comprising only the left hippocampus (3D models, $33\times46\times48$ voxels) or a selected 2D slice ($30\times36$ voxels) comprising part of the left hippocampus and amygdala -- both of these structures are highly affected by neurodegeneration in the early stages of AD \citep{jack2005brain, Knafo:2012}.

\subsection{Hyperparameters}
\label{HP_optimization}
Model hyperparameters were optimized as previously described in \citet{bruningk2021back}. The 2D model consists of three convolutional layers ($4\times4\times4$) and one fully connected layer. Convolutional layer inputs are padded such that the outputs have the same height and width as the inputs. Every convolutional layer is followed by a batch normalization layer, an activation layer (ReLU), and a max pooling layer ($2\times2$). The first two convolutional layers have 8 filters and the last convolutional layer has 16 filters. Dropout ($p=0.5$) is applied to the flattened outputs of the last convolutional layer. L2 regularization ($\lambda=0.001$) was applied to the weights of all layers. HPs associated with adversarial/interpretability aware training were optimized with grid search to maximize adversarial robustness.

The architecture of the 3D model was equivalent to that the 2D model. 2D convolutional and pooling layers were replaced with the corresponding 3D implementations. The learning rate, batch size, kernel sizes, and hyperparameters related to L2 regularization, dropout, and adversarial/interpretability aware training were kept constant. However, it is likely that the parameters relating to the adversarial attack and the interpretation discrepancy measure need further optimization.

Models were trained for a maximum of 3000 epochs on NVIDIA TITAN RTX, 24~GiB RAM GPUs using the Adam optimizer and a learning rate of 0.0003. Weights were initialized using the He method \citep{he2015delving}. The models return the raw (pre-softmax) score of the sample for each of the two classes in the model. 

\newpage
\section{Supplementary Results}\label{apd:suppl}
\begin{table}[h]
% \scriptsize%\footnotesize 
\tiny
\floatconts
  {tab:normal}%
  {\caption{Comparing normal and combined baseline performance on benign 1.5T and 3T examples. * Not evaluated.}}%
  {\begin{tabular}{@{}lc@{\phantom{.}}c@{\phantom{.}}c@{\phantom{.}}c@{\phantom{.}}c|c@{\phantom{.}}c@{\phantom{.}}c@{\phantom{.}}c@{\phantom{.}}c}\toprule
    Test &\multicolumn{5}{c|}{3T} &\multicolumn{5}{c}{1.5T}\\
    Train & ACC & TPR & TNR & AUC & APS & ACC & TPR & TNR & AUC & APS\\
    \midrule
    2D-1.5T & 0.82 & 0.78 & 0.80 & 0.90 & 0.86 & 0.75 & \textbf{0.79} & 0.70 & \textbf{0.85} & \textbf{0.87}\\
    2D-3T & \textbf{0.88} & 0.79 & \textbf{0.90} & \textbf{0.94} & \textbf{0.90} & 0.73 & 0.72 & 0.72 & 0.82 & 0.85\\
    2D-3T/1.5T & 0.87 & \textbf{0.83} & 0.88 & 0.93 & \textbf{0.90} & \textbf{0.77} & 0.78 & \textbf{0.76} & \textbf{0.85} & \textbf{0.87} \\
    \midrule
    3D-1.5T & * & * & * & * & * & \textbf{0.87} & \textbf{0.90} & \textbf{0.83} & \textbf{0.93} & \textbf{0.93}\\
    3D-3T & \textbf{0.88} & 0.75 & \textbf{0.94} & \textbf{0.95} & \textbf{0.92} & 0.44 & 0.20 & 0.73 & 0.50 & 0.60\\
    3D-3T/1.5T & \textbf{0.88} & \textbf{0.77} & 0.92 & 0.94 & 0.90 & 0.82 & 0.84 & 0.81 & 0.91 & 0.92 \\
    \bottomrule
    \end{tabular}}
\end{table}

\begin{figure}[htbp]
\floatconts
  {fig:suppl_1_out}
  {%
    \addtocounter{figure}{-1}
    \subfigure[AUC]{\label{fig:AUC_1_out}%
      \includegraphics[width=\linewidth]{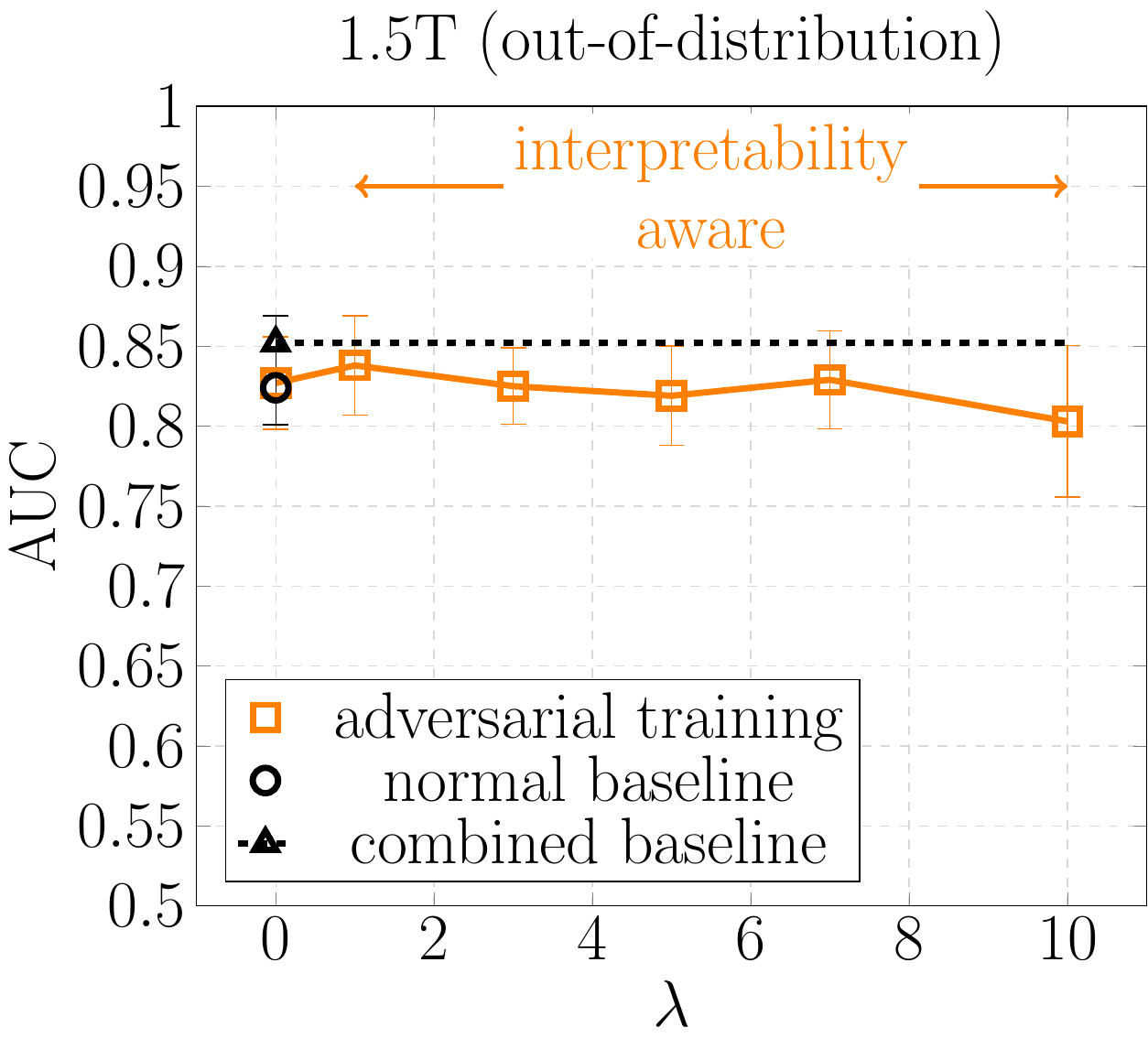}}%
    \qquad
    \subfigure[APS]{\label{fig:APS_1_out}%
      \includegraphics[width=\linewidth]{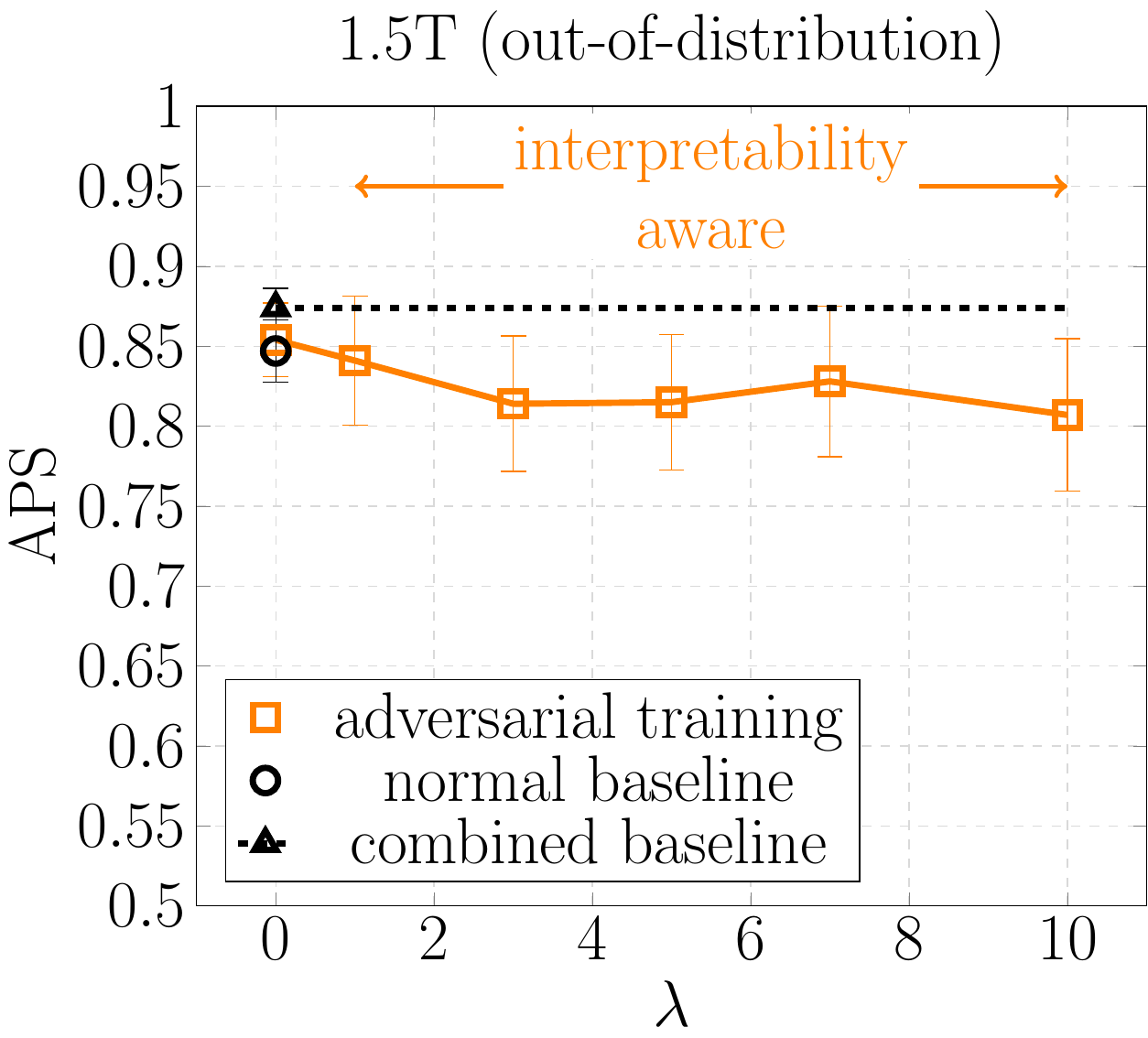}}
  }
  {\caption{\textbf{2D OOD} performance metrics shown as a function of $\lambda$. Performance is evaluated on 2D OOD (1.5T) data. The adversarial model ($\lambda = 0$) and the interpretability aware models ($\lambda > 0$) were trained using $\pmb{\epsilon = 0.001}$. The `normal' and `combined' baseline indicated by a circle and a triangle are trained on undisturbed data.}}
\end{figure}

\begin{figure*}[htbp]
\addtocounter{figure}{1}
\floatconts
  {fig:suppl_1_in}
  {%
    \addtocounter{figure}{-1}
    \subfigure[TPR]{\label{fig:TPR_1_in}%
      \includegraphics[width=0.3\linewidth]{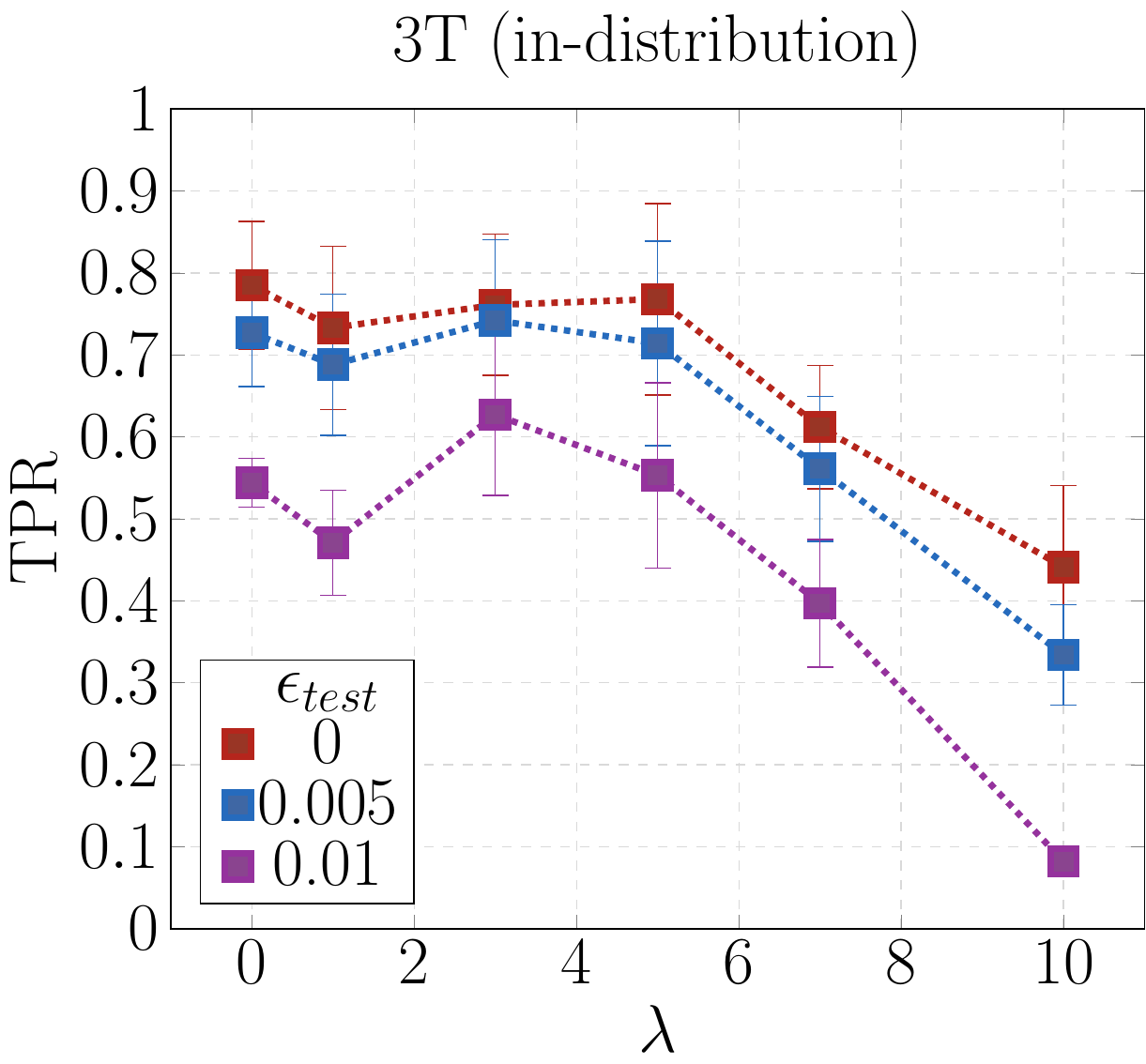}}%
    \qquad
    \subfigure[TNR]{\label{fig:TNR_1_in}%
      \includegraphics[width=0.3\linewidth]{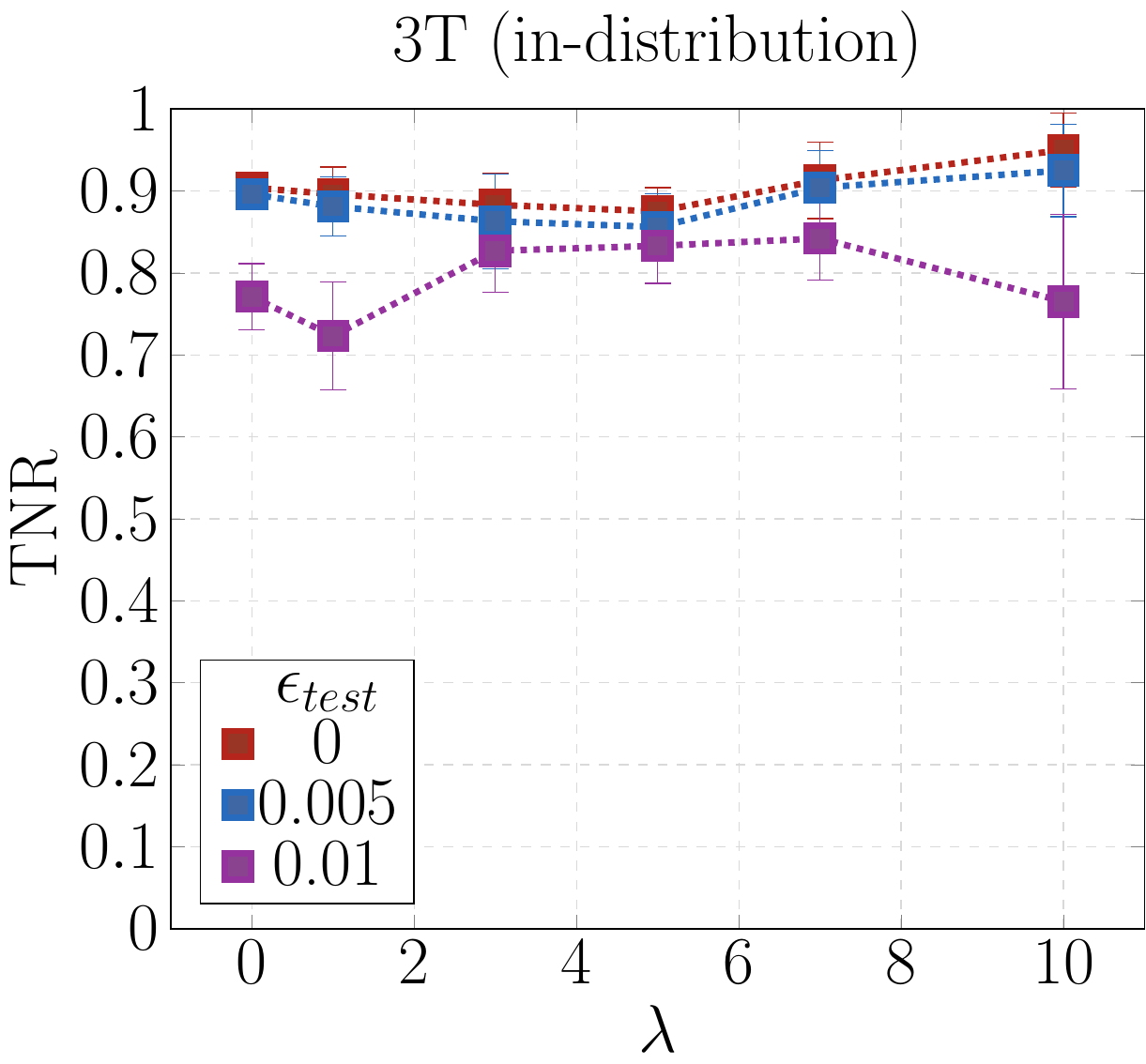}}
    \qquad
    \subfigure[ACC]{\label{fig:ACC_1_in}%
      \includegraphics[width=0.3\linewidth]{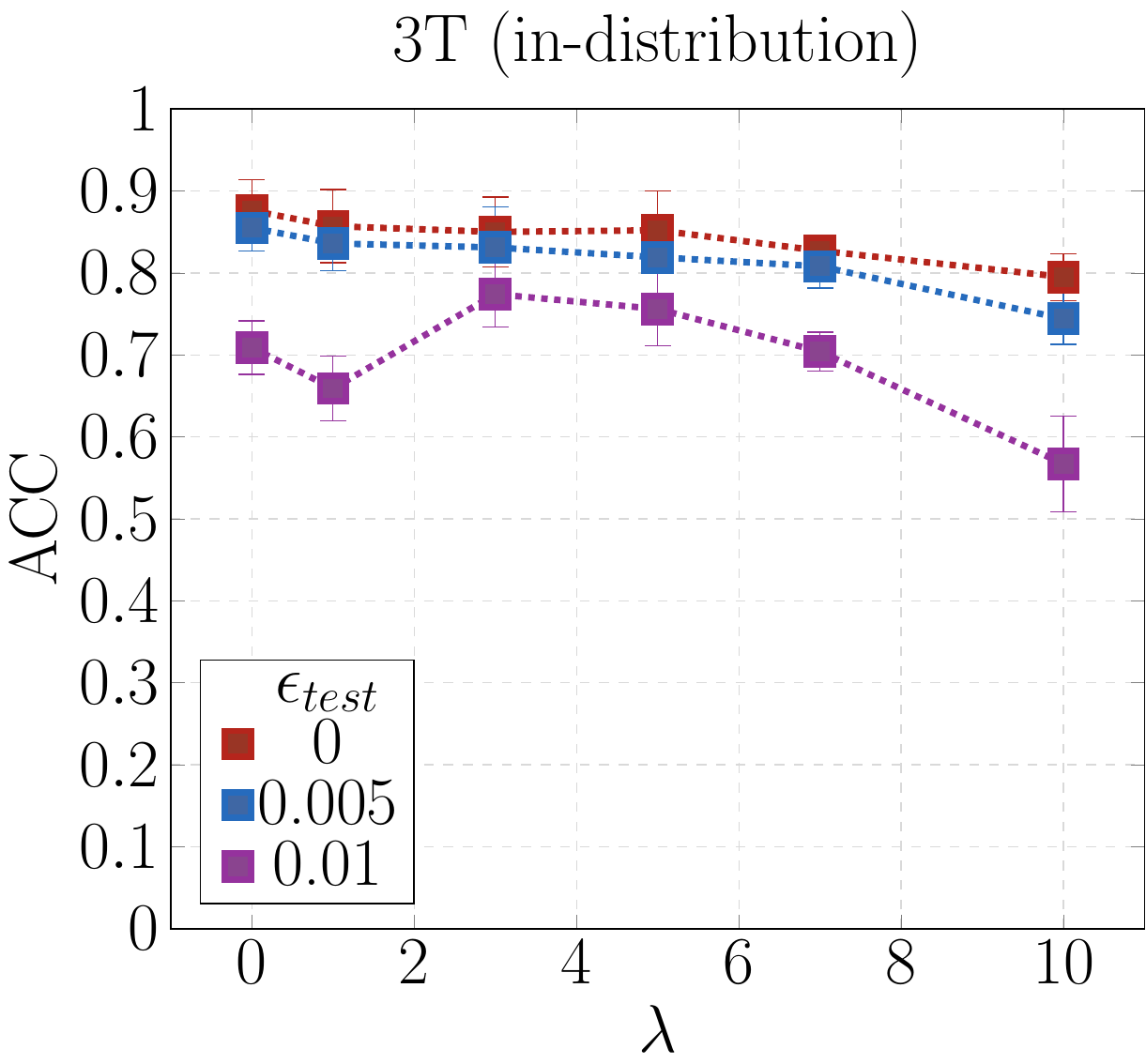}}
    \qquad
    \subfigure[AUC]{\label{fig:AUC_1_in}%
      \includegraphics[width=0.3\linewidth]{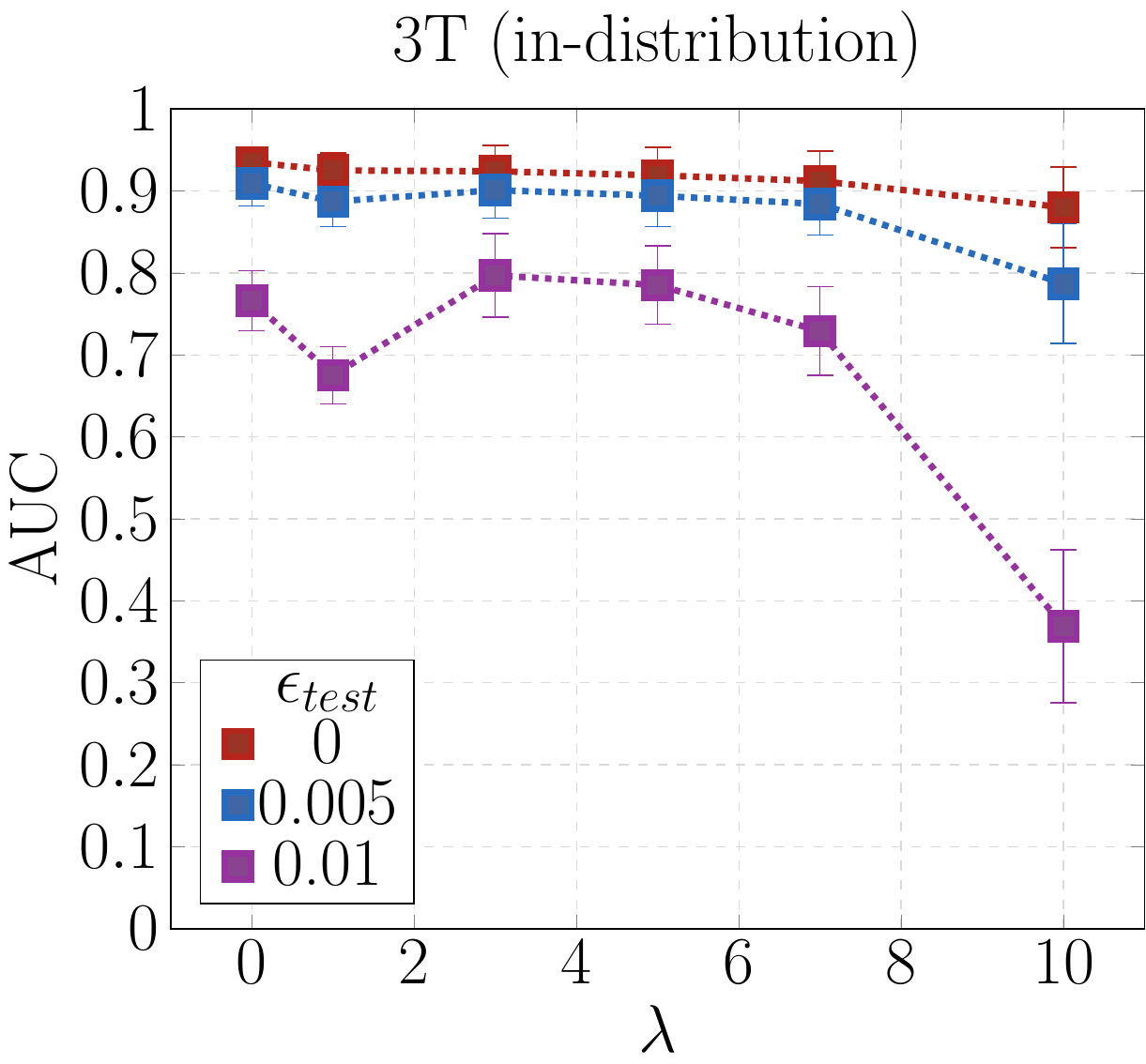}}%
    \qquad
    \subfigure[APS]{\label{fig:APS_1_in}%
      \includegraphics[width=0.3\linewidth]{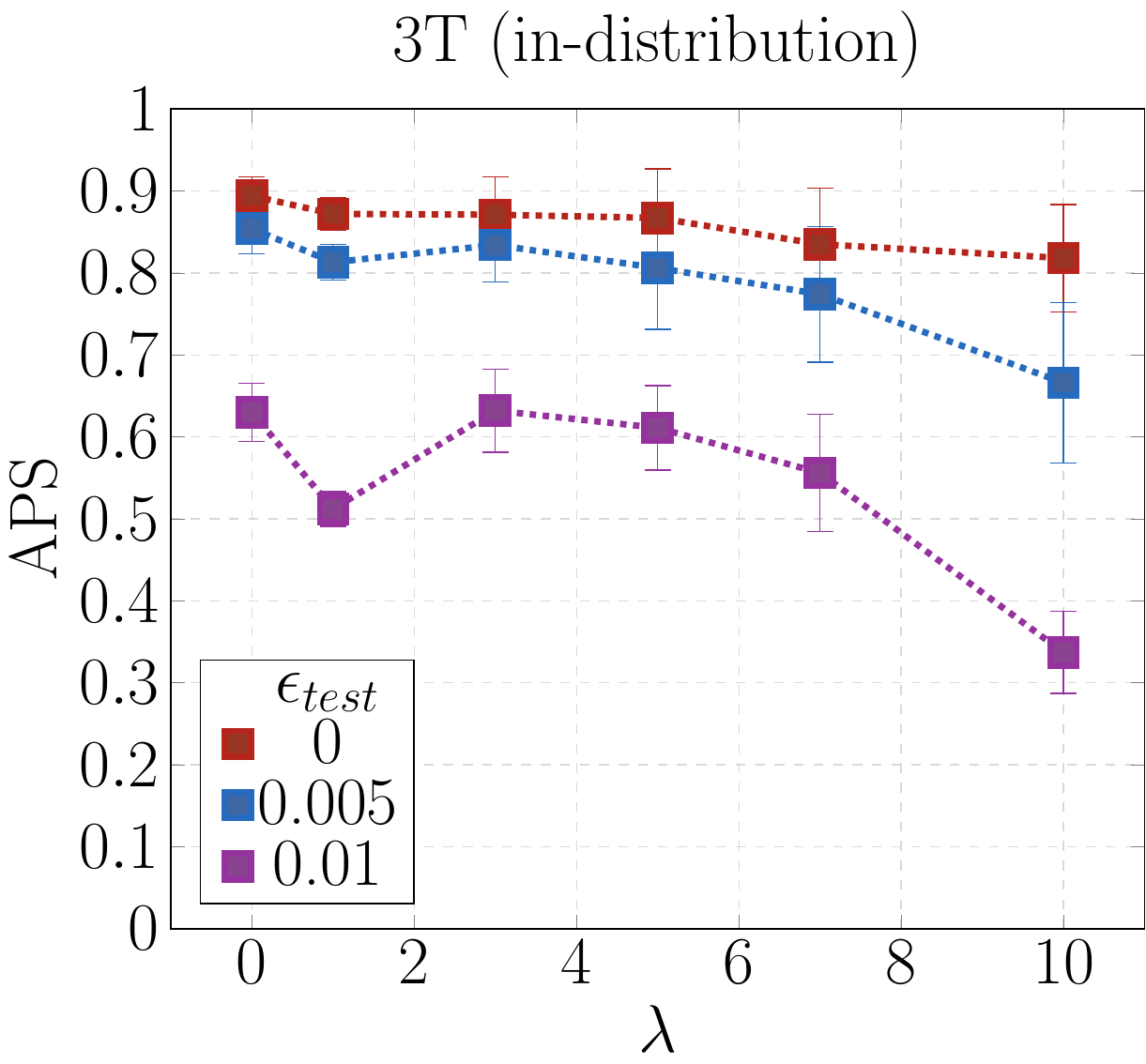}}
  }
  {\caption{\textbf{2D within-distribution} performance metrics shown as a function of $\lambda$. Performance is evaluated on benign ($\epsilon_{test}=0$) and disturbed ($\epsilon_{test}>0$) 2D within-distribution (3T) images. All models were trained using $\pmb{\epsilon = 0.001}$. As expected, performance generally degraded with increasing disturbance of the test images. Beyond $\lambda = 5$, performance on both benign and perturbed samples degrades notably, i.e. strong emphasis on class activation map agreement comes at the cost of a higher misclassification rate. Comparing the model trained under the adversarial regime with the models trained under the interpretability aware regime, we observe that interpretability awareness conveyed a performance advantage on adversarially perturbed 3T images for an \textit{optimal} choice of $\lambda$, $\lambda \in \{3, 5\}$. The advantage becomes more pronounced as the adversary becomes stronger.}}
\end{figure*}

\begin{figure*}[htbp]
\addtocounter{figure}{1}
\floatconts
  {fig:metrics_5_out}
  {%
    \addtocounter{figure}{-1}
    \subfigure[TPR]{\label{fig:TPR_5_out}%
      \includegraphics[width=0.3\linewidth]{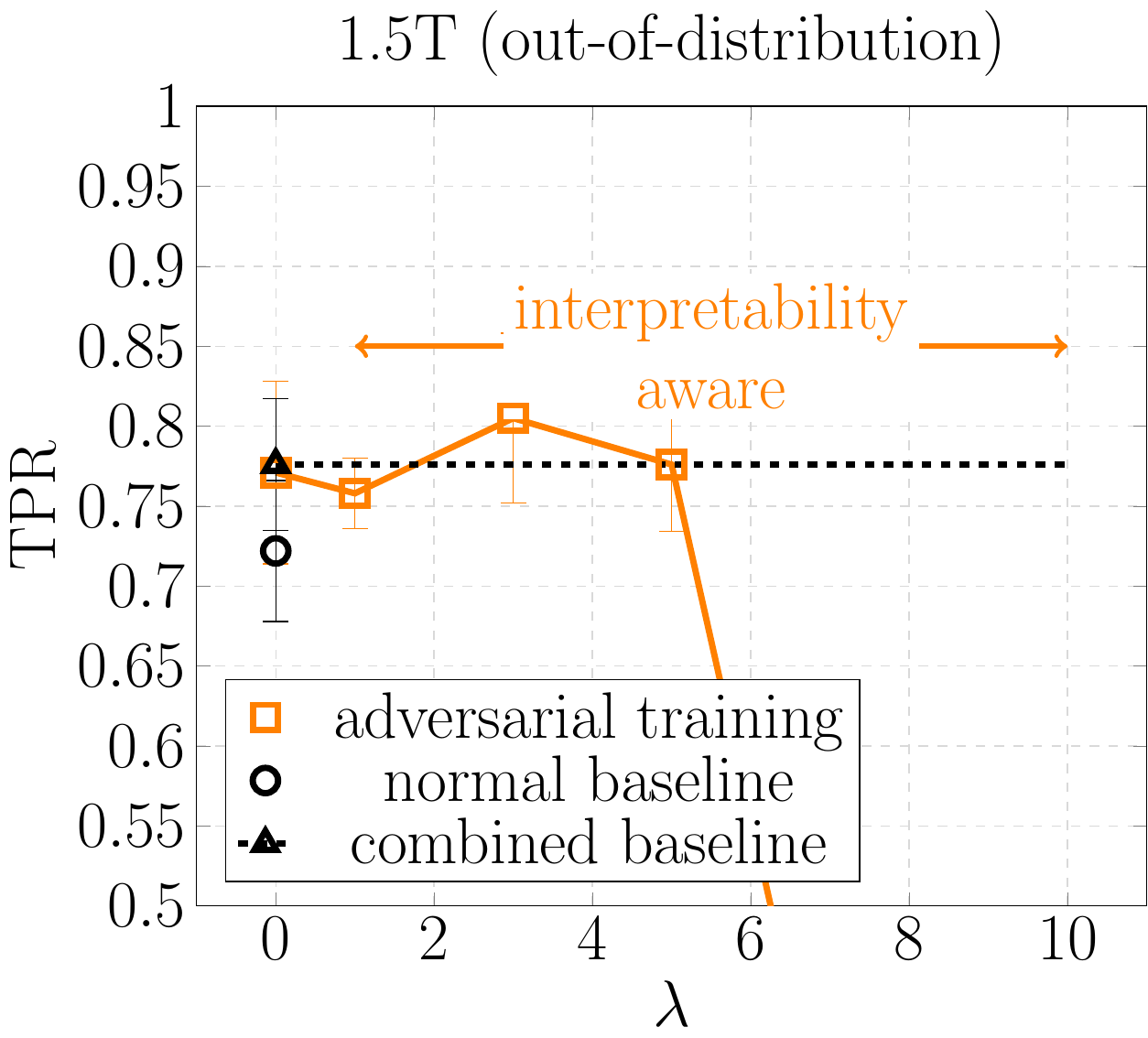}}%
    \qquad
    \subfigure[TNR]{\label{fig:TNR_5_out}%
      \includegraphics[width=0.3\linewidth]{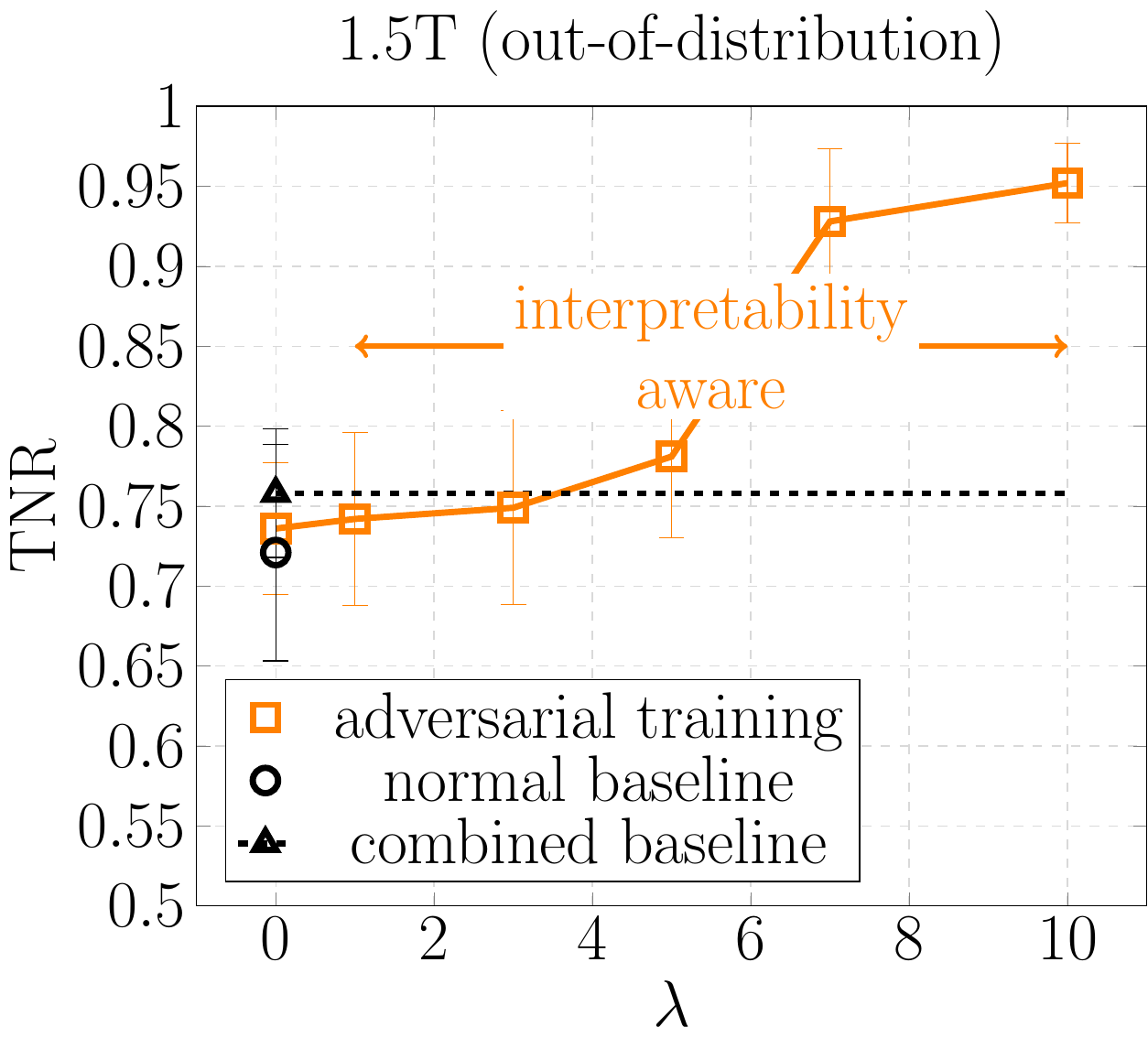}}
    \qquad
    \subfigure[ACC]{\label{fig:ACC_5_out}%
      \includegraphics[width=0.3\linewidth]{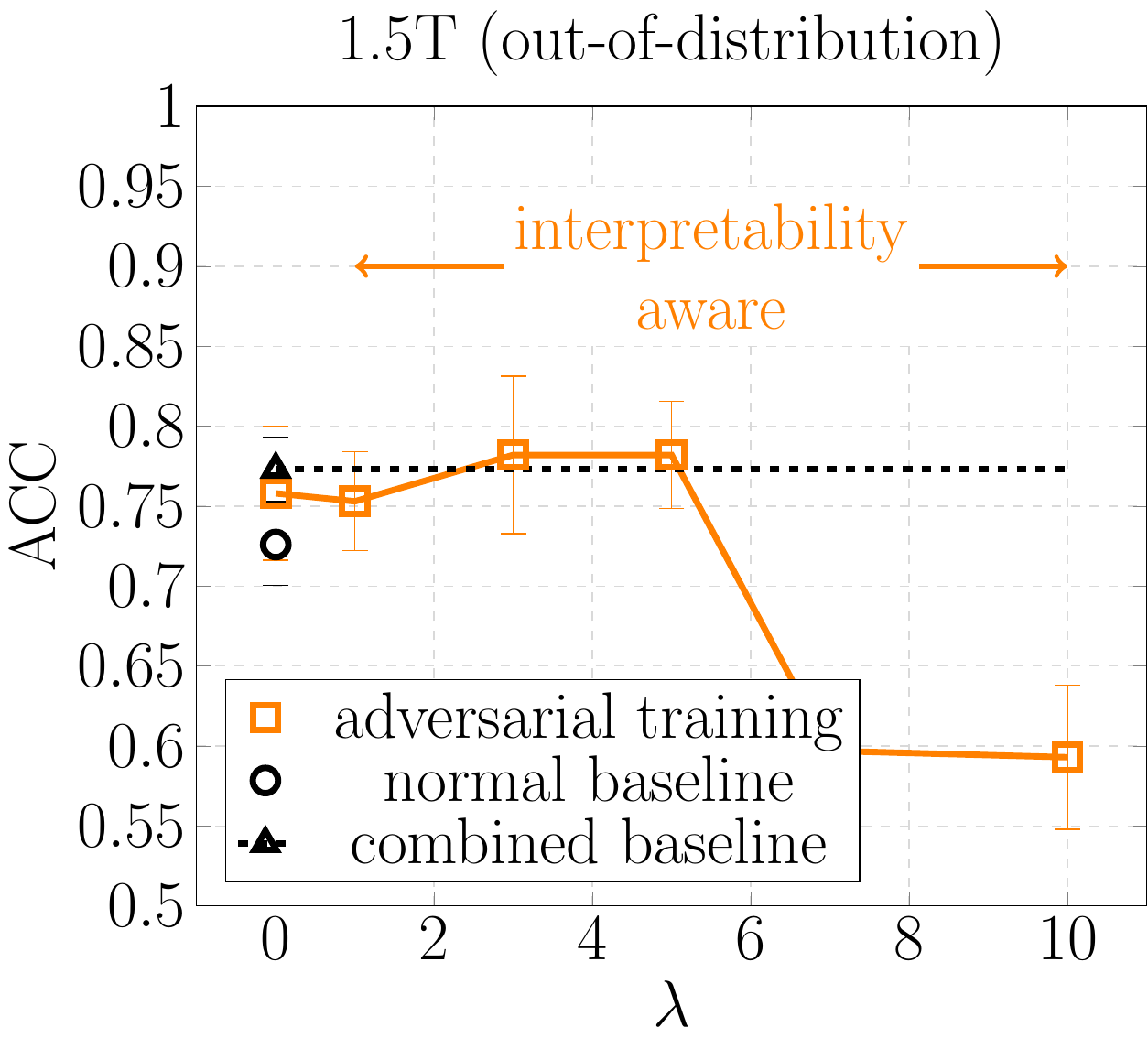}}
    \qquad
    \subfigure[AUC]{\label{fig:AUC_5_out}%
      \includegraphics[width=0.3\linewidth]{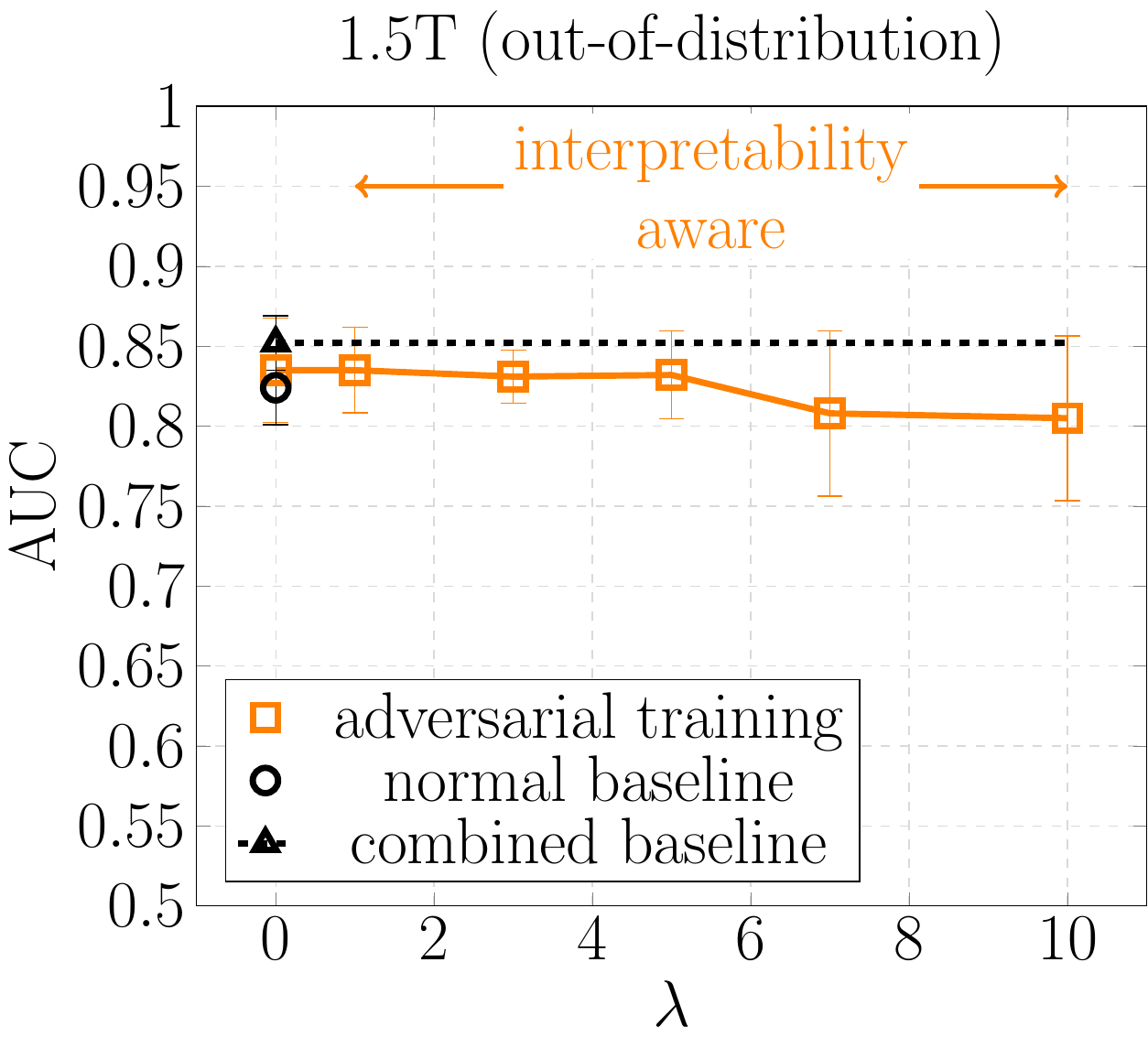}}%
    \qquad
    \subfigure[APS]{\label{fig:APS_5_out}%
      \includegraphics[width=0.3\linewidth]{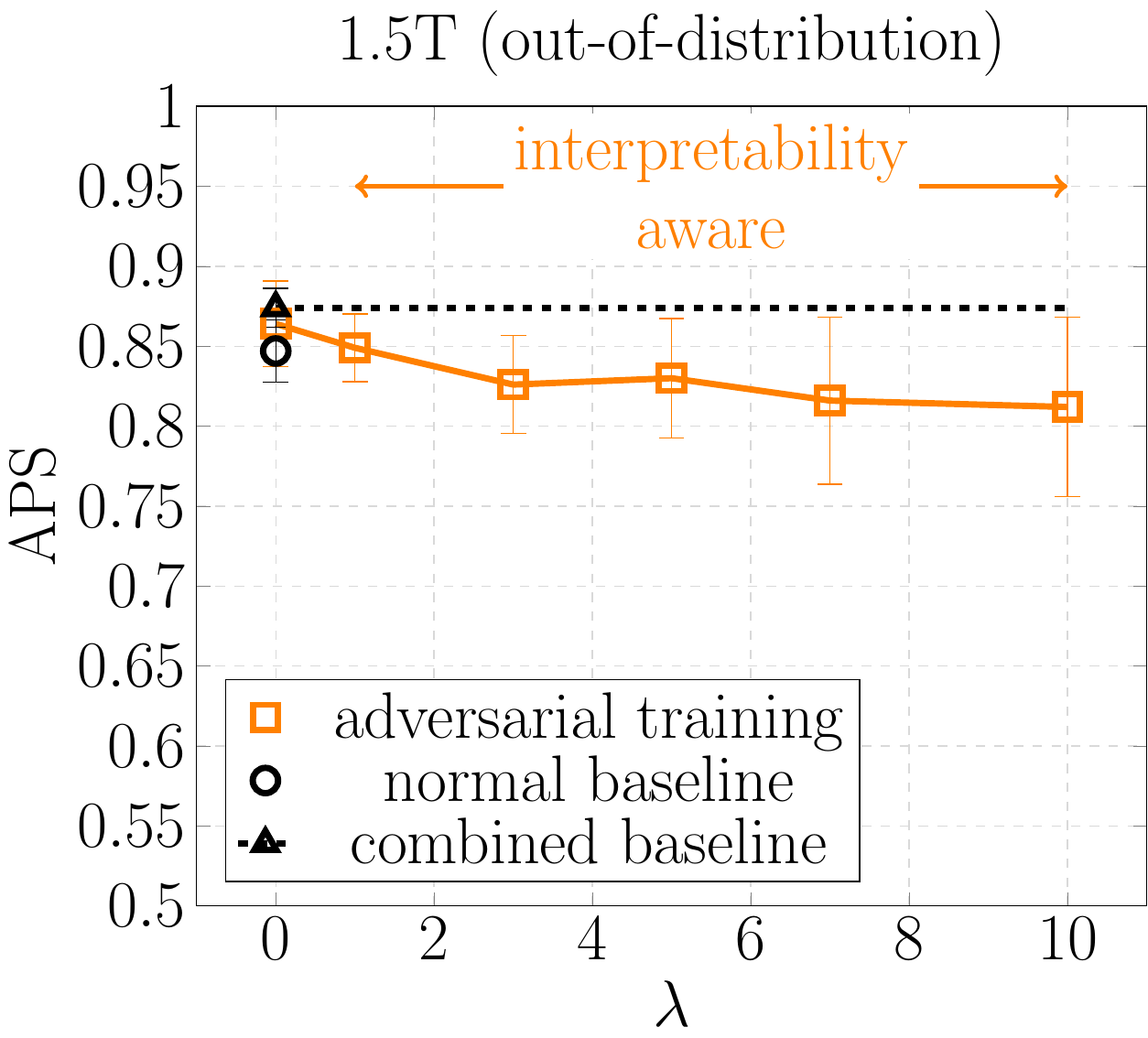}}
  }
  {\caption{\textbf{2D OOD} performance metrics shown as a function of $\lambda$. Performance is evaluated on 2D OOD (1.5T) data. Adversarial and interpretability aware models were trained using $\pmb{\epsilon = 0.005}$. The `normal' and `combined' baseline indicated by a circle and a triangle are trained on undisturbed data. Models failed to train when $\epsilon_{train}$ was further increased to 0.01. The $\lambda = 3$ and $\lambda = 5$ interpretability aware models performed better on the unperturbed 1.5T data compared to the adversarial model and the normal baseline. Performance of the best interpretability aware models is on par with that of the combined baseline. % The figure indicates that models trained with stronger adversarial examples ($\epsilon_{train} = 0.005$) also show comparable benign performance (e.g. $\lambda$ = 5, TPR = 0.731, TNR = 0.900, ACC = 0.858), suggesting that this higher limit on the strength of the adversarial attack does not affect performance on benign samples. 
  }}
\end{figure*}

\begin{figure*}[htbp]
\addtocounter{figure}{1}
\floatconts
  {fig:metrics_5_in}
  {%
    \addtocounter{figure}{-1}
    \subfigure[TPR]{\label{fig:TPR_5_in}%
      \includegraphics[width=0.3\linewidth]{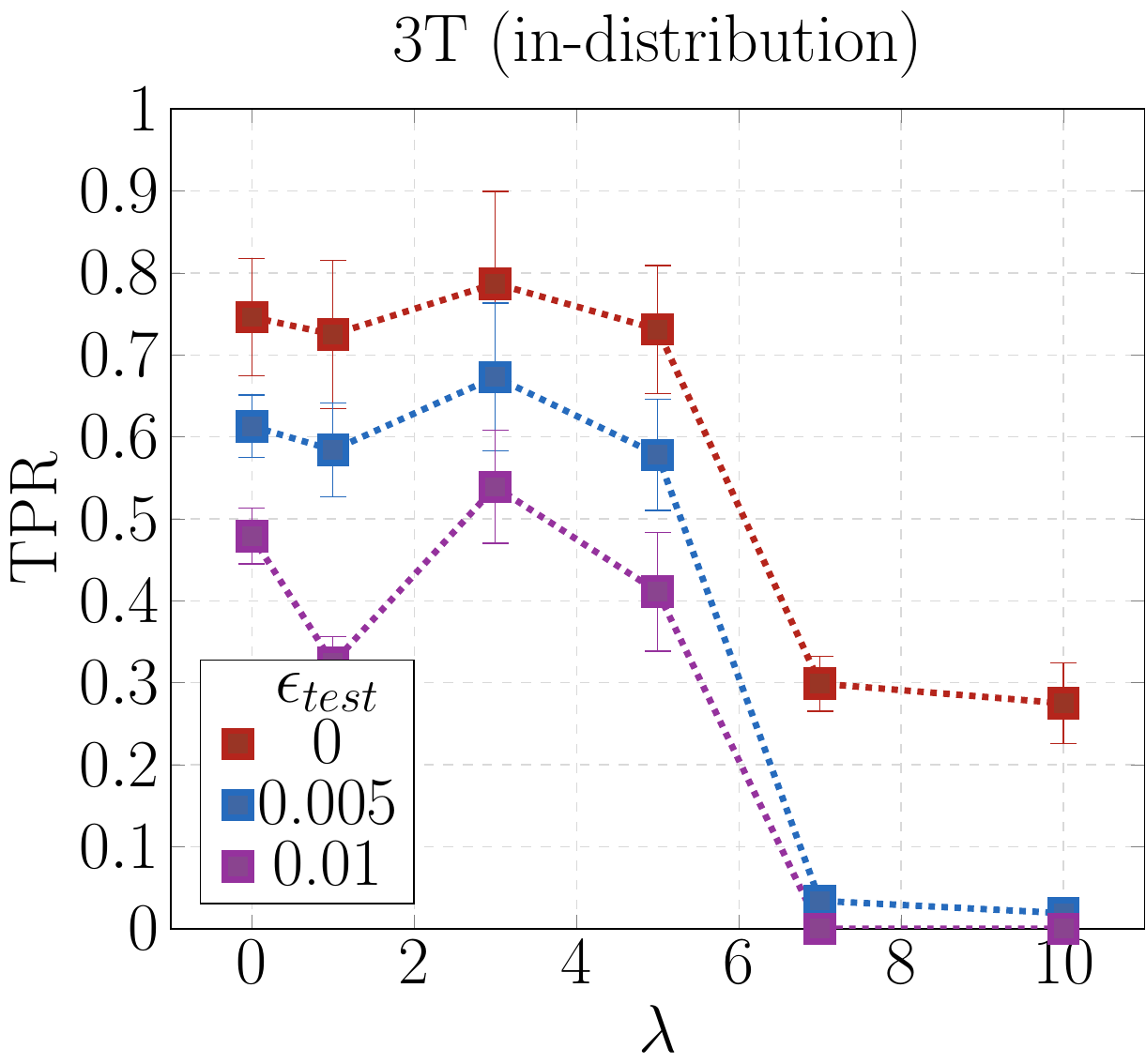}}%
    \qquad
    \subfigure[TNR]{\label{fig:TNR_5_in}%
      \includegraphics[width=0.3\linewidth]{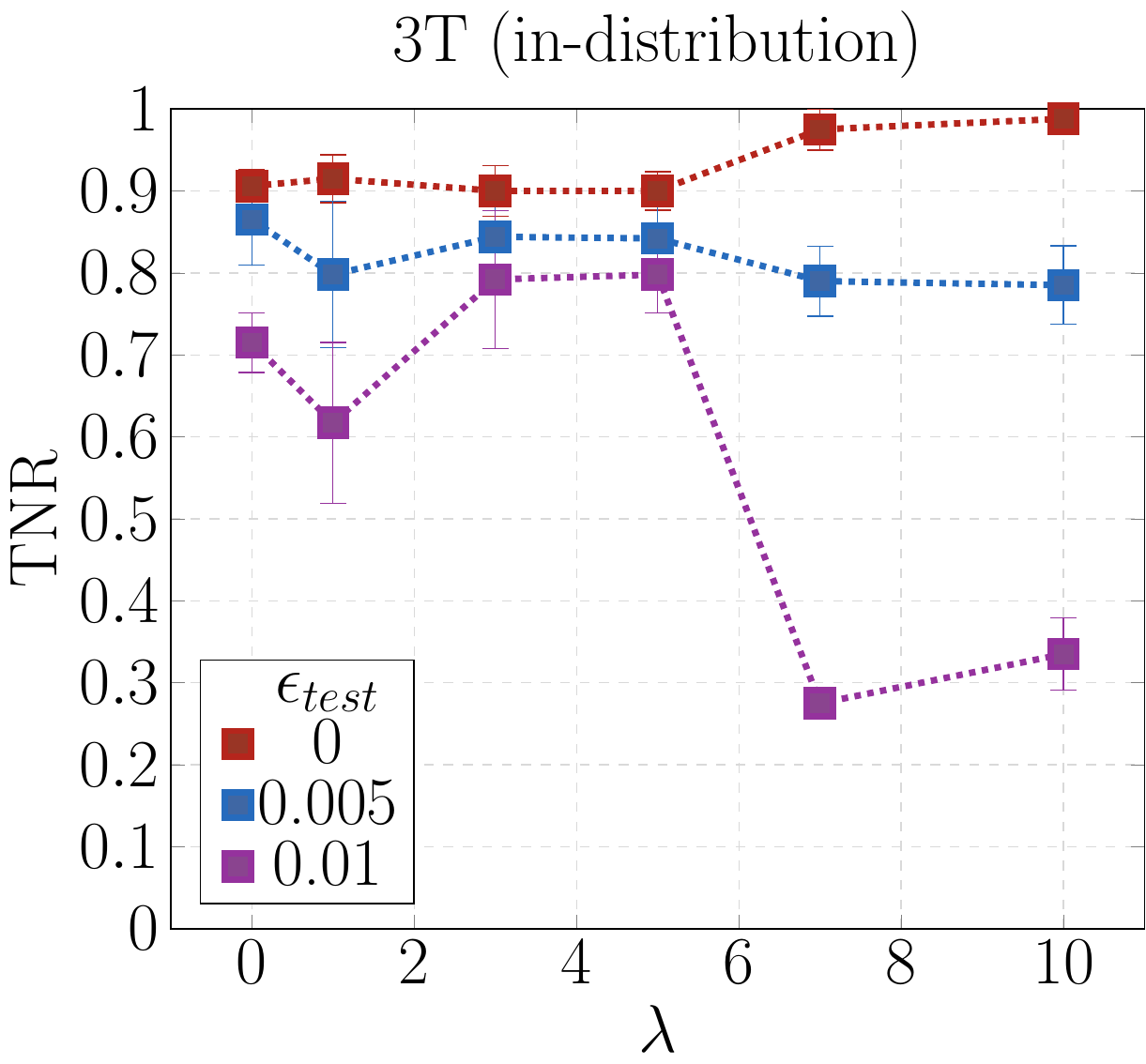}}
    \qquad
    \subfigure[ACC]{\label{fig:ACC_5_in}%
      \includegraphics[width=0.3\linewidth]{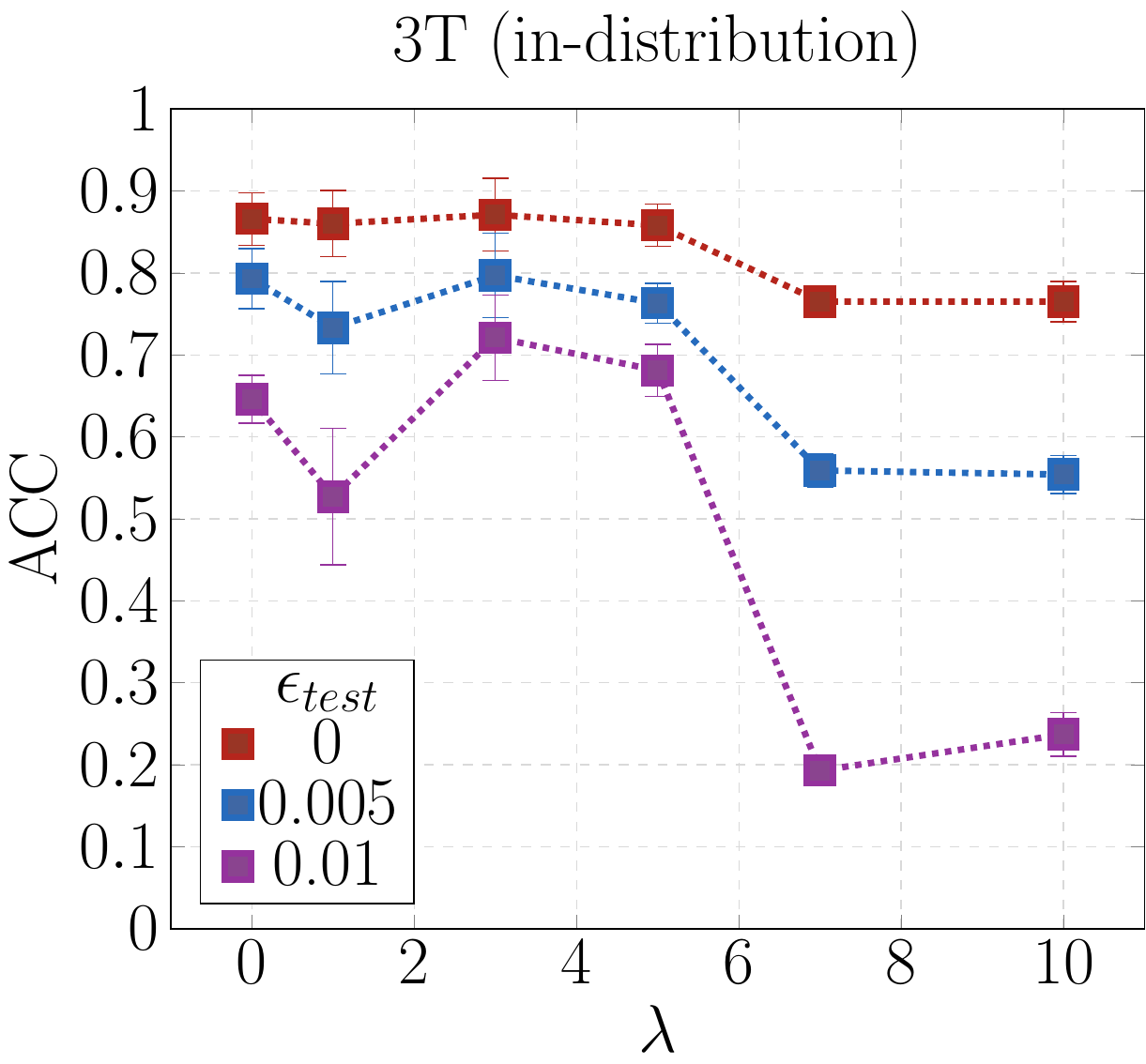}}
    \qquad
    \subfigure[AUC]{\label{fig:AUC_5_in}%
      \includegraphics[width=0.3\linewidth]{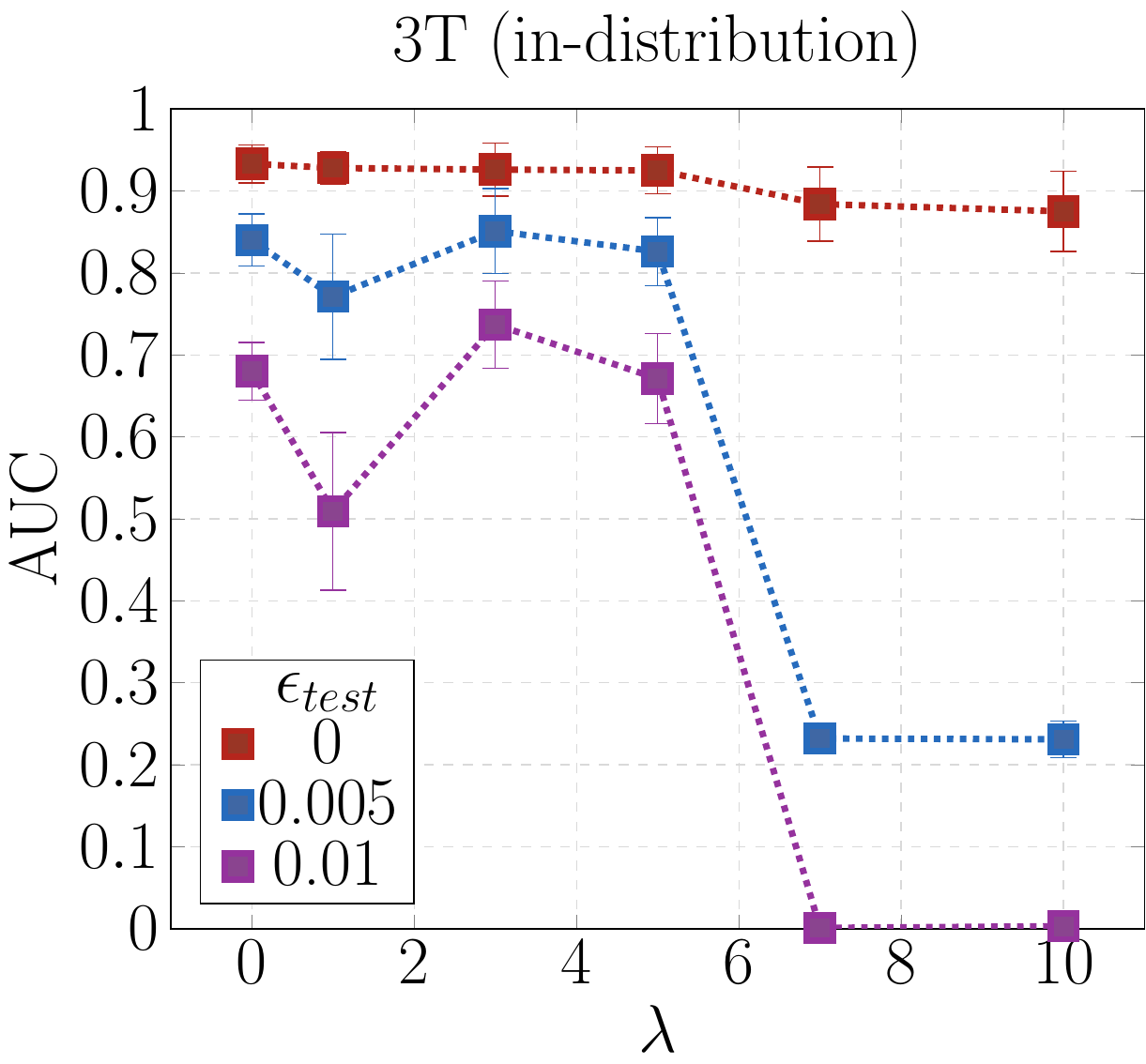}}%
    \qquad
    \subfigure[APS]{\label{fig:APS_5_in}%
      \includegraphics[width=0.3\linewidth]{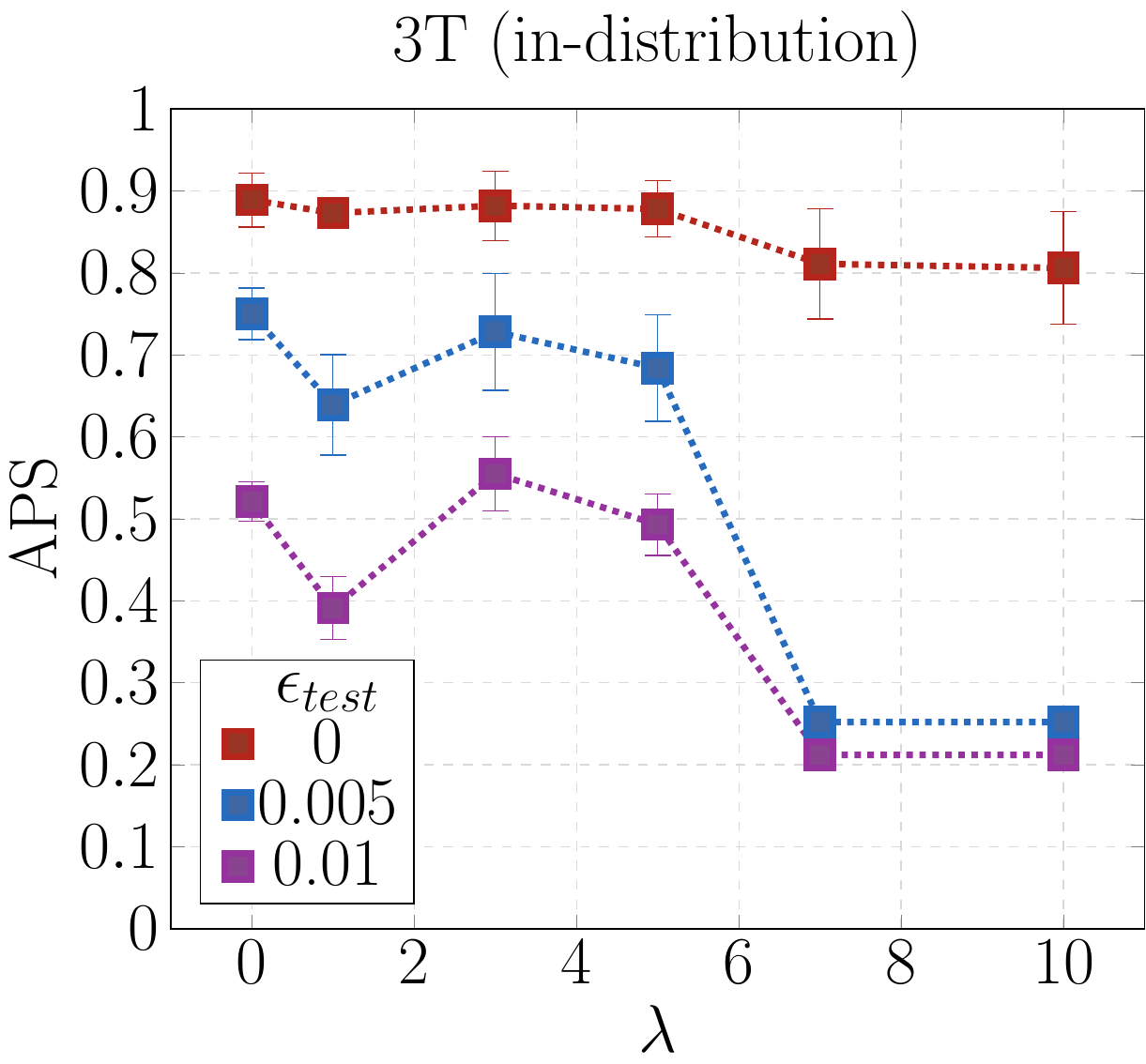}}
  }
  {\caption{\textbf{2D within-distribution} performance metrics shown as a function of $\lambda$. Performance is evaluated on benign ($\epsilon_{test}=0$) and disturbed ($\epsilon_{test}>0$) 2D within-distribution (3T) images. All models were trained using $\pmb{\epsilon = 0.005}$. Models failed to train when $\epsilon_{train}$ was further increased to 0.01. Interpretability aware models with $\lambda \leq 5$ perform similarly to adversarially trained models on unperturbed 3T test images ($\epsilon_{test} = 0$). Performance degrades when $\lambda > 5$. Given a strong adversary ($\epsilon_{test} = 0.01$), the $\lambda = 3$ interpretability aware model outperforms the adversarial model both in terms of TPR and TNR. % The figure indicates that models trained with stronger adversarial examples ($\epsilon_{train} = 0.005$) also show comparable benign performance (e.g. $\lambda$ = 5, TPR = 0.731, TNR = 0.900, ACC = 0.858), suggesting that this higher limit on the strength of the adversarial attack does not affect performance on benign samples. 
  }}
\end{figure*}

\begin{figure*}[htbp]
\addtocounter{figure}{1}
\floatconts
  {fig:3Dmetrics_out}
  {%
    \addtocounter{figure}{-1}
    \subfigure[TPR]{\label{fig:3D_TPR_1_out}%
      \includegraphics[width=0.3\linewidth]{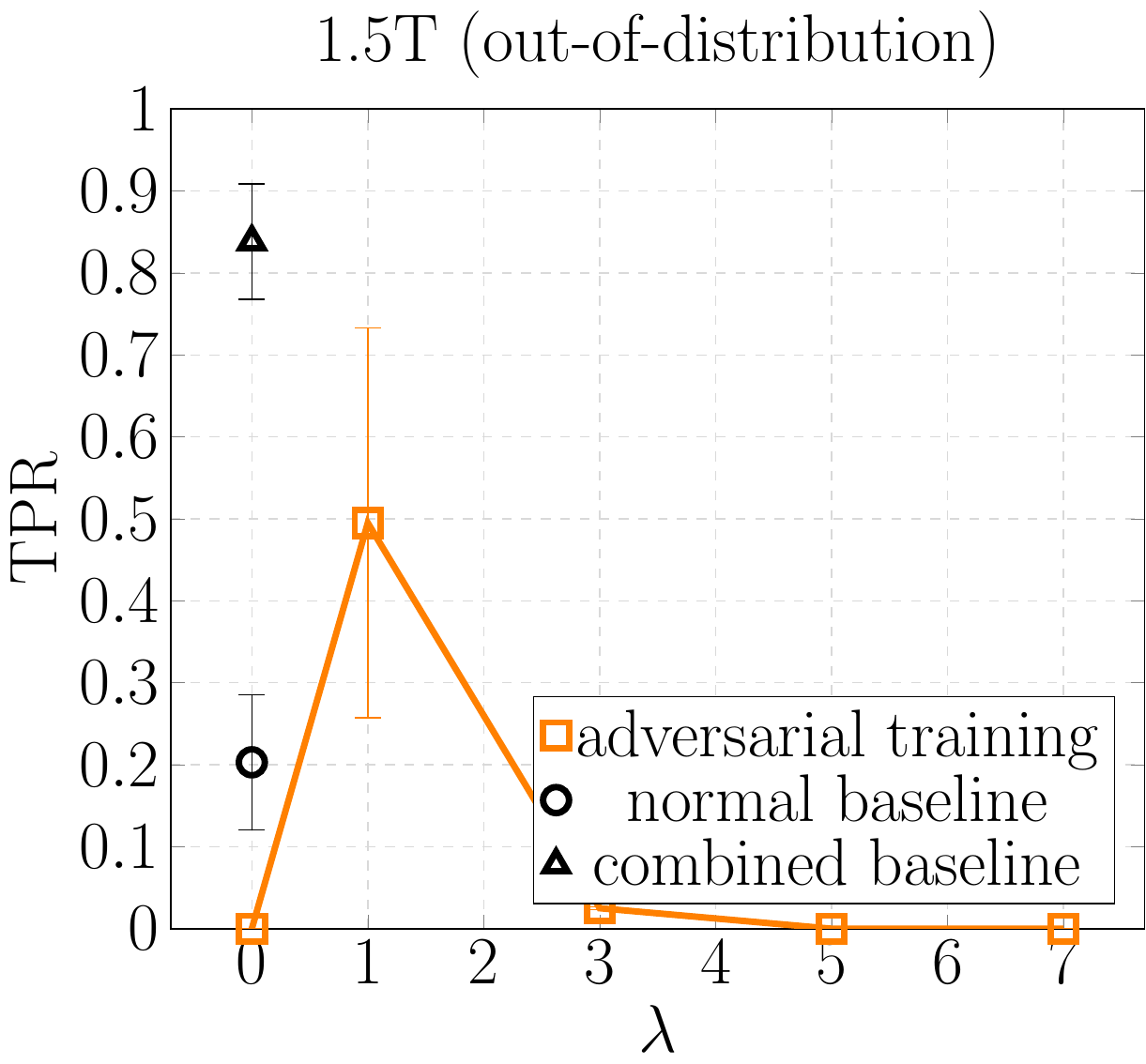}}%
    \qquad
    \subfigure[TNR]{\label{fig:3D_TNR_1_out}%
      \includegraphics[width=0.3\linewidth]{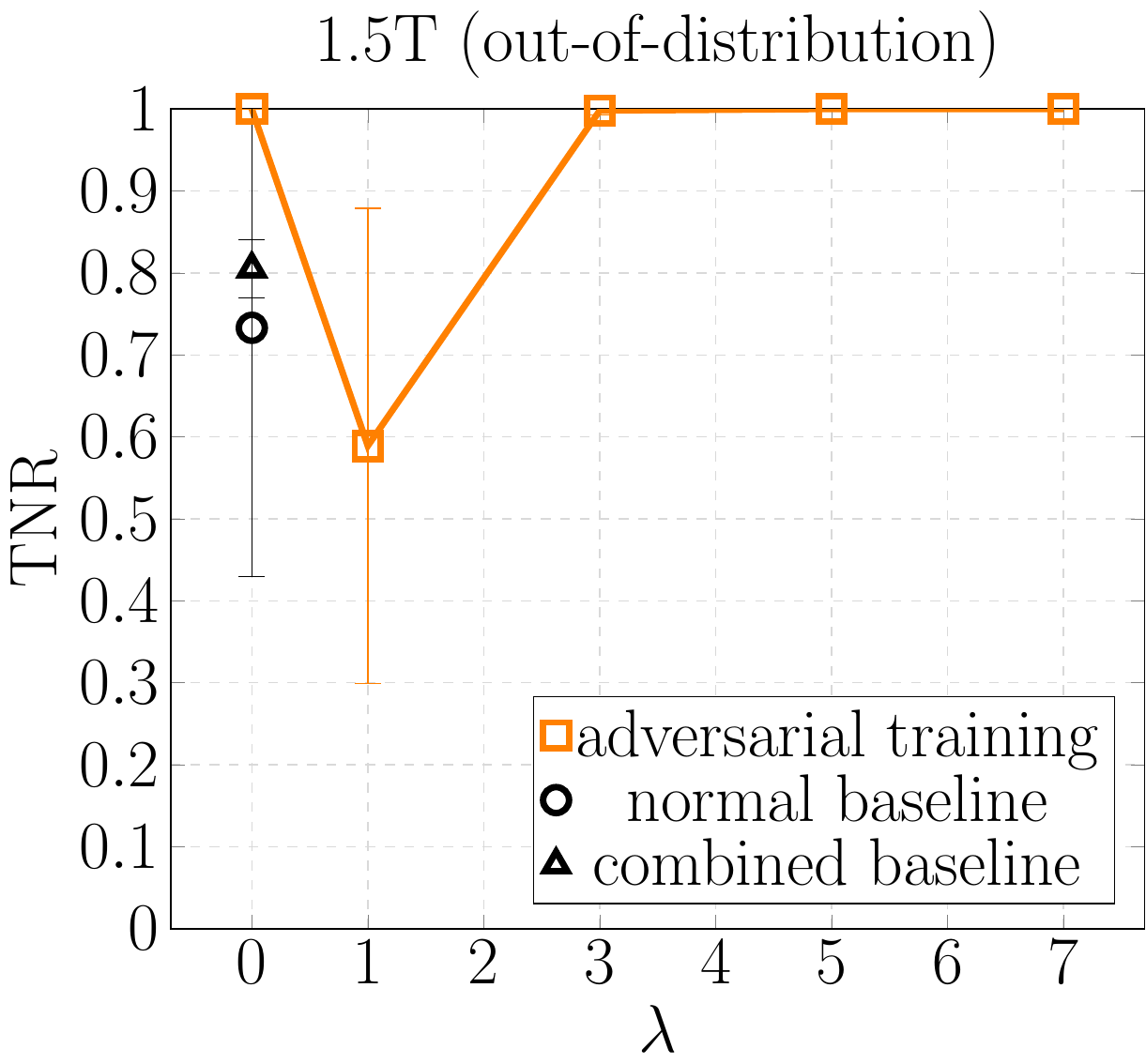}}
    \qquad
    \subfigure[ACC]{\label{fig:3D_ACC_1_out}%
      \includegraphics[width=0.3\linewidth]{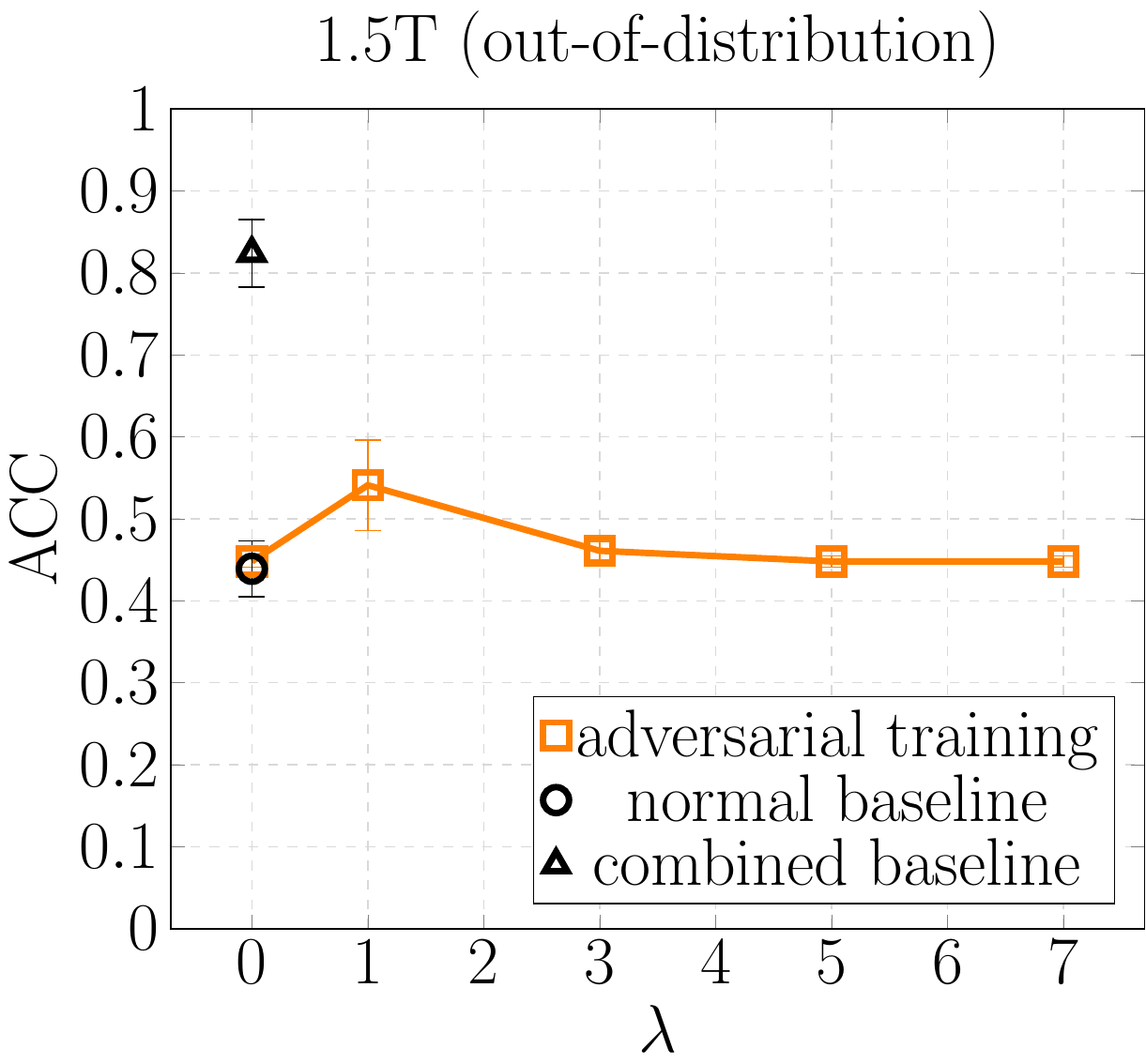}}
    \qquad
    \subfigure[AUC]{\label{fig:3D_AUC_1_out}%
      \includegraphics[width=0.3\linewidth]{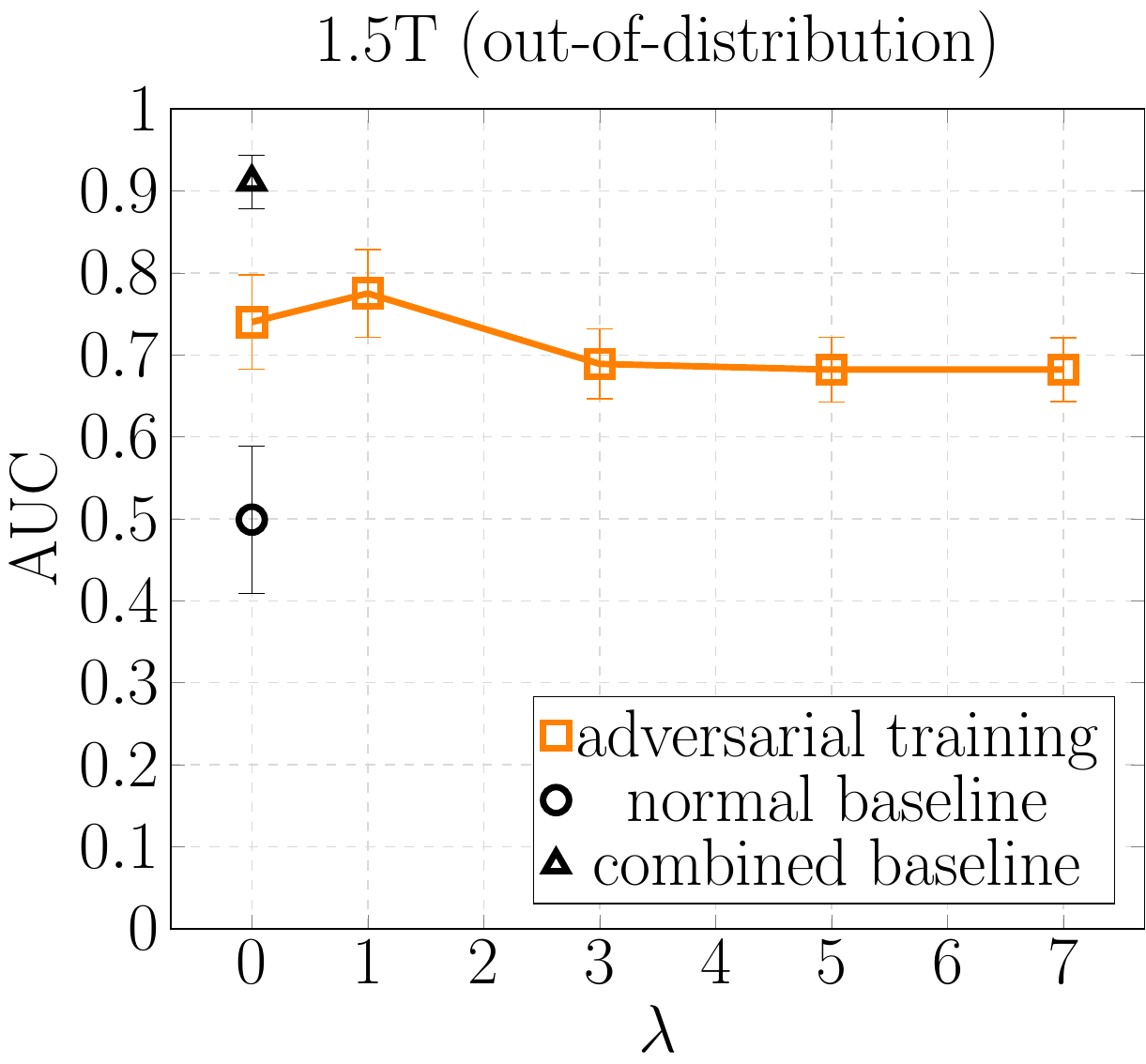}}
    \qquad
    \subfigure[APS]{\label{fig:3D_APS_1_out}%
      \includegraphics[width=0.3\linewidth]{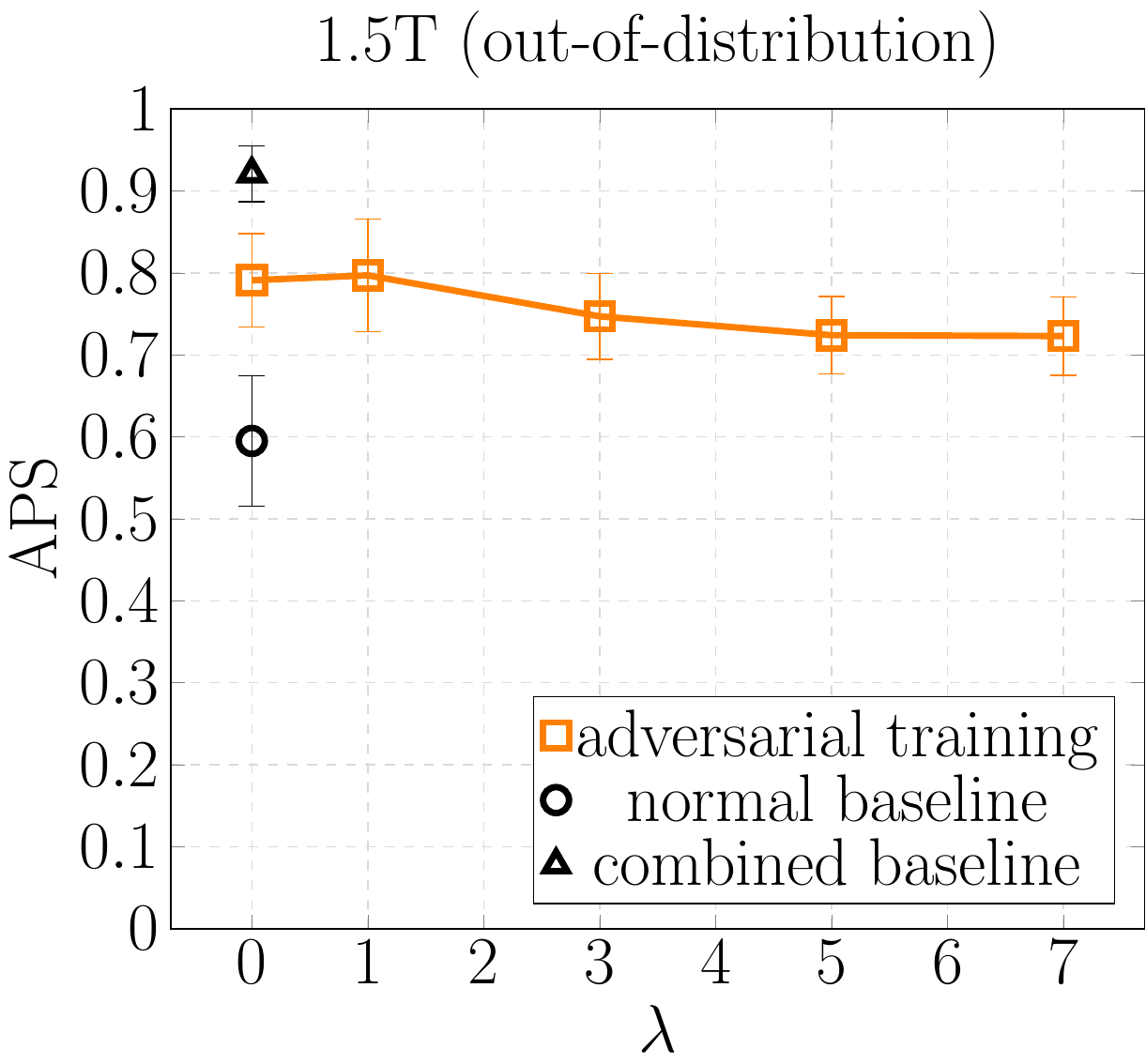}}
  }
  {\caption{\textbf{3D OOD} performance metrics shown as a function of $\lambda$. Performance is evaluated on 3D OOD (1.5T) data. Adversarial and interpretability aware models were trained using $\pmb{\epsilon = 0.001}$. The `normal' and `combined' baseline indicated by a circle and a triangle are trained on undisturbed data. Interpretability hyperparameters are the same as in the 2D setting and it is likely that a separate hyperparameter optimization procedure would benefit performance.
  }}
\end{figure*}

\begin{figure*}[htbp]
\addtocounter{figure}{1}
\floatconts
  {fig:3Dmetrics_in}
  {%
    \addtocounter{figure}{-1}
    \subfigure[TPR]{\label{fig:3D_TPR_1_in}%
      \includegraphics[width=0.3\linewidth]{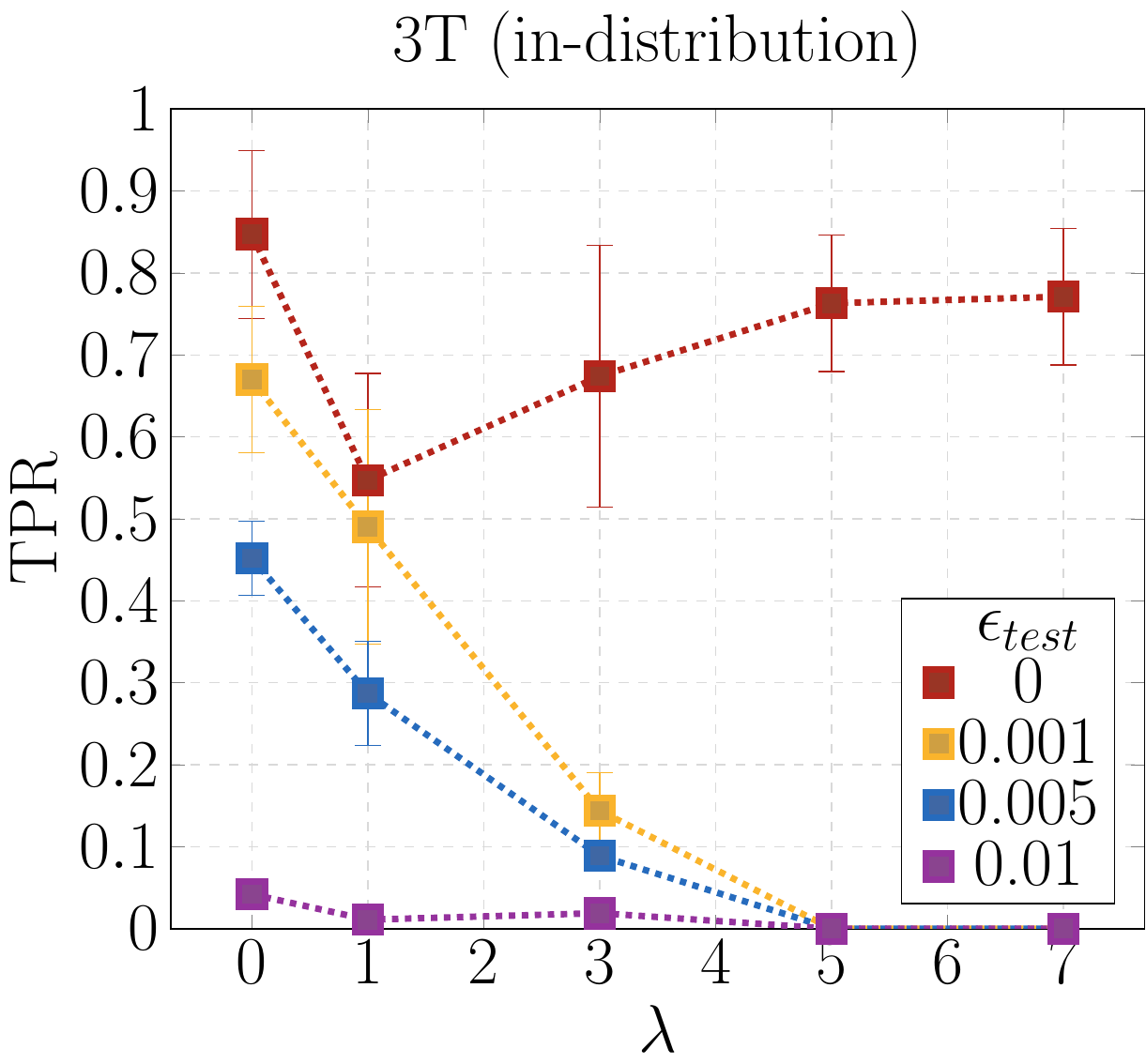}}
    \qquad
    \subfigure[TNR]{\label{fig:3D_TNR_1_in}%
      \includegraphics[width=0.3\linewidth]{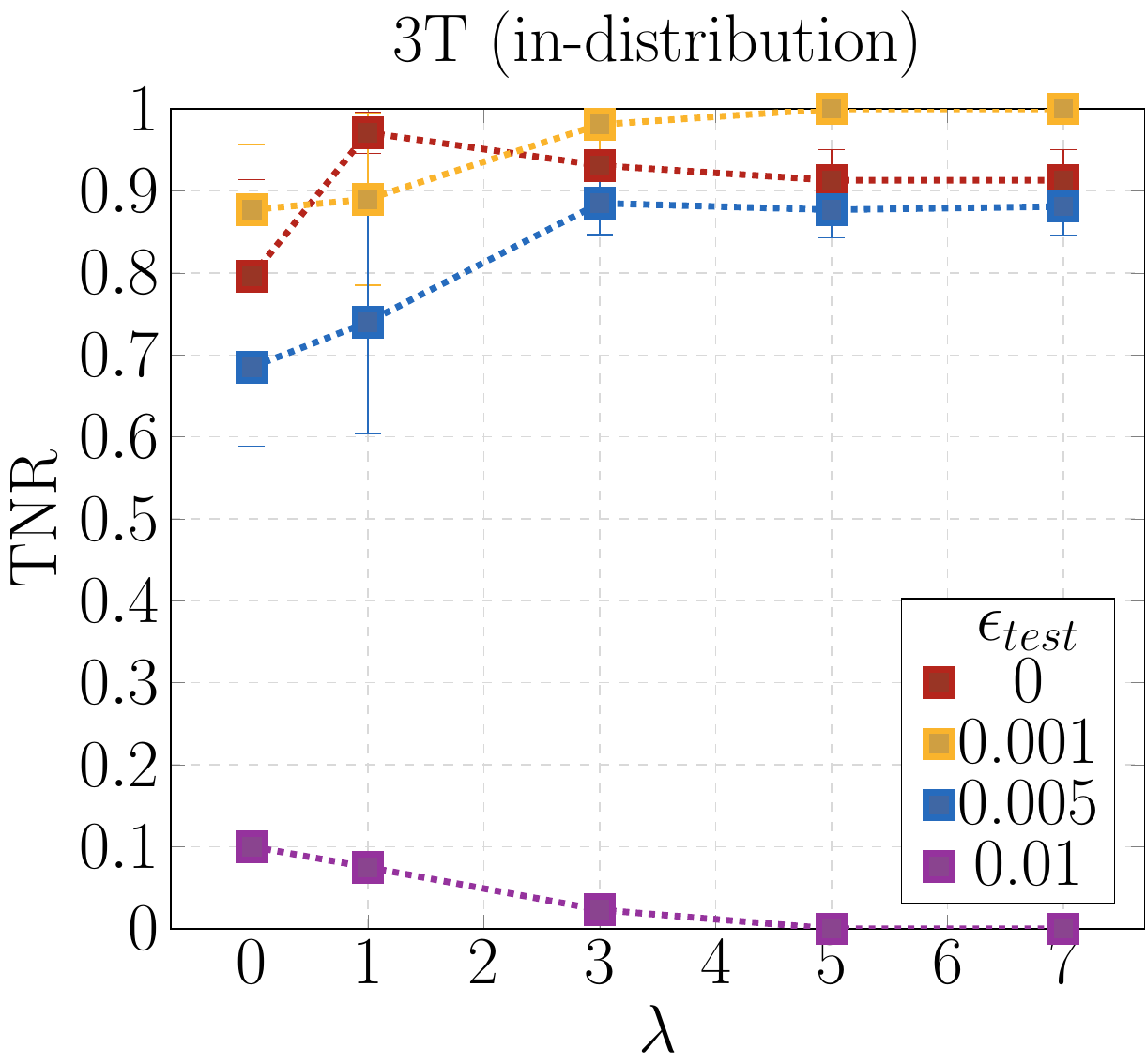}}
    \qquad
    \subfigure[ACC]{\label{fig:3D_ACC_1_in}%
      \includegraphics[width=0.3\linewidth]{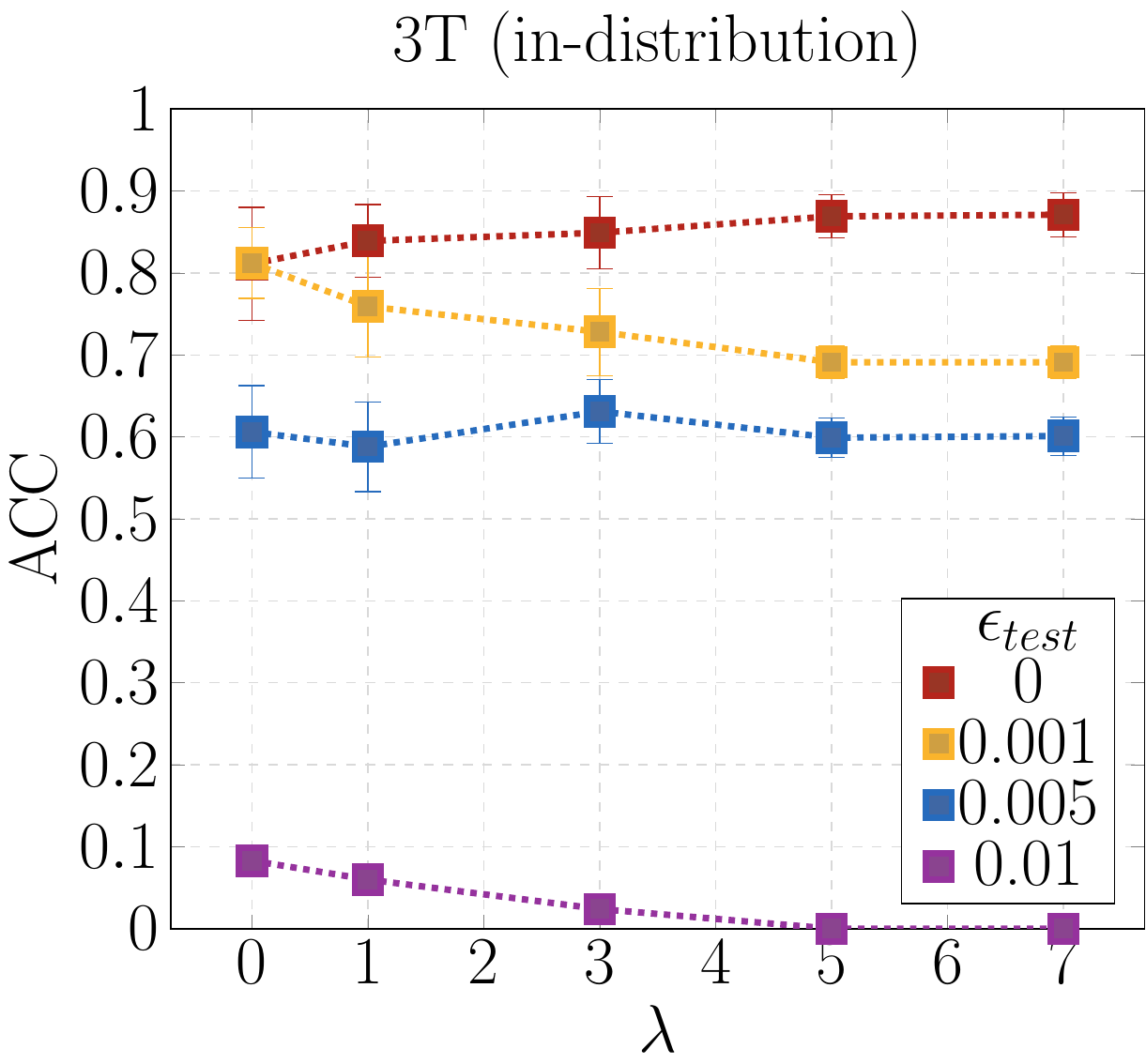}}
    \qquad
    \subfigure[AUC]{\label{fig:3D_AUC_1_in}%
      \includegraphics[width=0.3\linewidth]{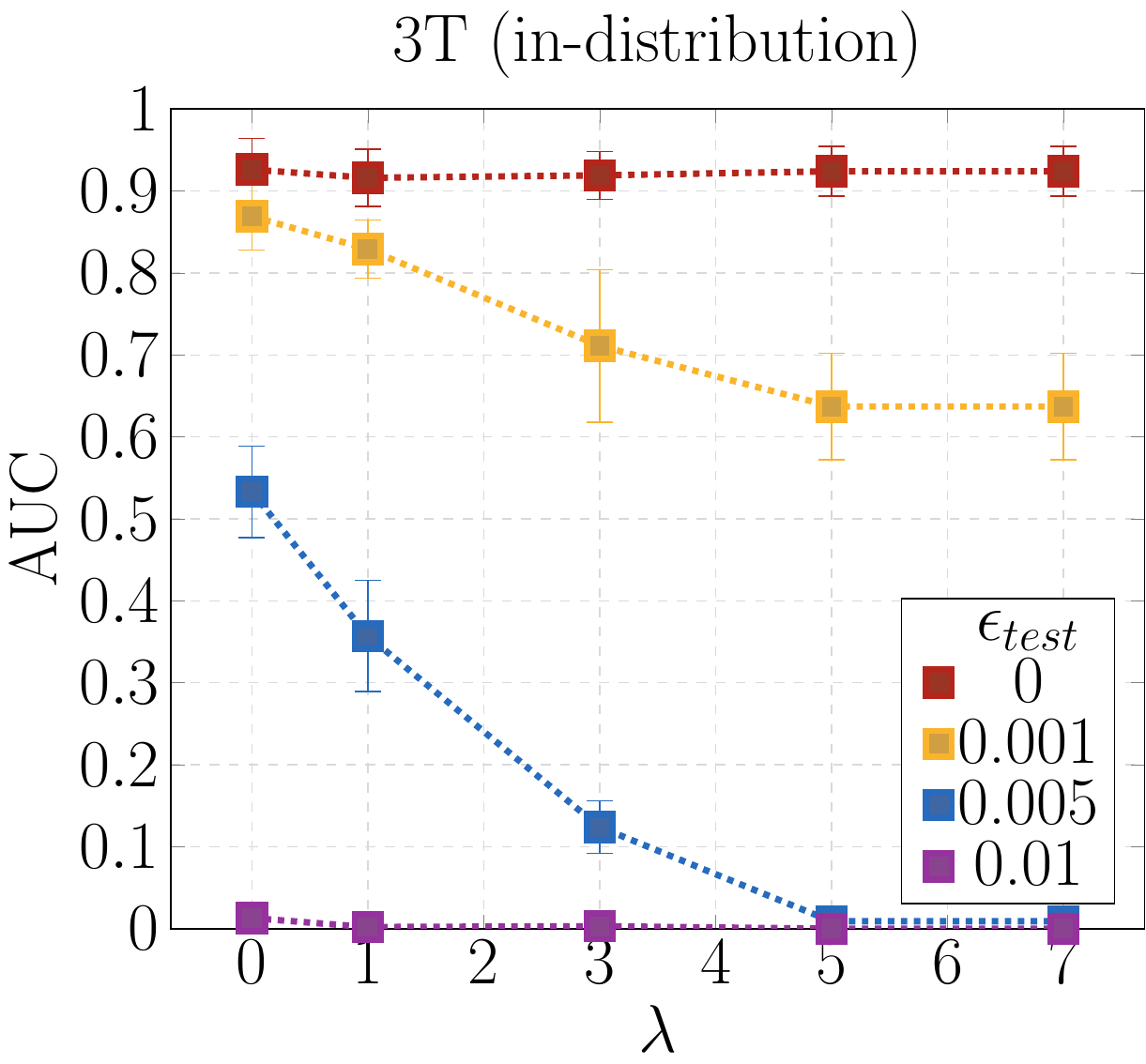}}
    \qquad
    \subfigure[APS]{\label{fig:3D_APS_1_in}%
      \includegraphics[width=0.3\linewidth]{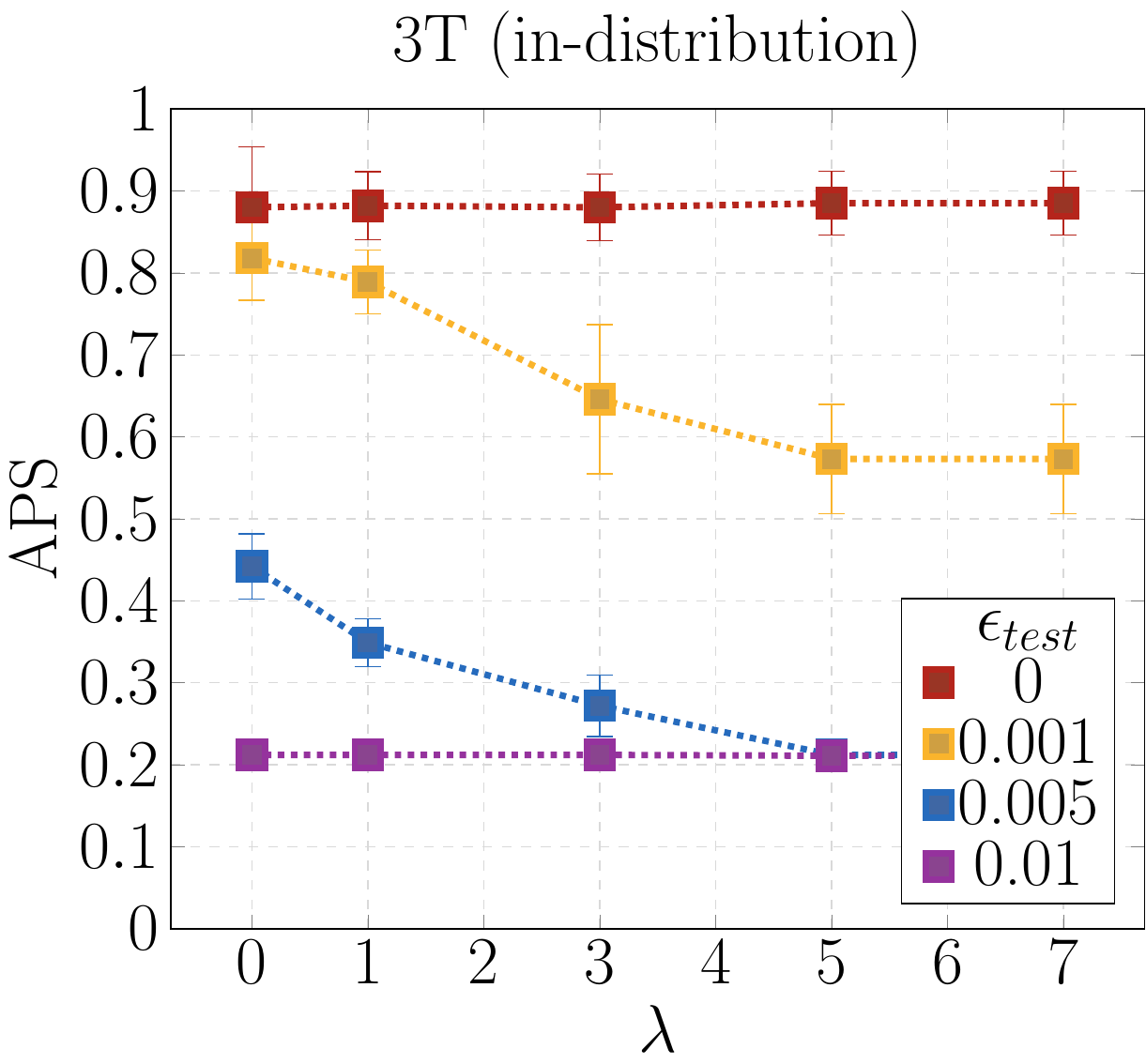}}
  }
  {\caption{\textbf{3D within-distribution} performance metrics shown as a function of $\lambda$. Performance is evaluated on benign ($\epsilon_{test}=0$) and disturbed ($\epsilon_{test}>0$) 3D within-distribution (3T) images. All models were trained using $\pmb{\epsilon = 0.001}$. Attack and interpretability hyperparameters are the same as in the 2D setting and it is likely that a separate hyperparameter optimization procedure would benefit performance.
  }}
\end{figure*}

\begin{figure*}[htbp]
\addtocounter{figure}{1}
\floatconts
  {fig:gradcam_tn_c0}
  {%
    \addtocounter{figure}{-1}
    \subfigure[CN subject: CN evidence (\textit{int})]{\label{fig:tn_c0_int}%
      \includegraphics[width=0.98\linewidth]{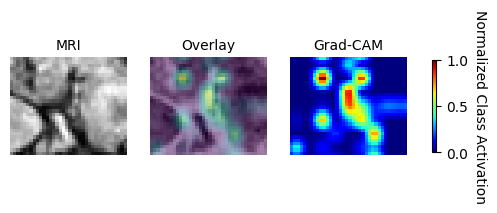}}
    \qquad
    \subfigure[CN subject: CN evidence (\textit{adv})]{\label{fig:tn_c0_adv}%
      \includegraphics[width=0.98\linewidth]{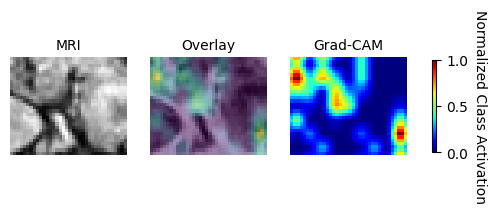}}
  }
{\caption{Representative saliency maps of an \textbf{OOD CN} example. Both the adversarial (\textit{adv}) and the interpretability aware (\textit{int}) model classified this example correctly. The maps highlight the CN evidence. The saliency map produced by the interpretability aware model highlights brain tissue surrounding the grooves, whereas the adversarial saliency map highlights both tissue and gaps, including features on the edge of the image. 
  }}
\end{figure*}

\begin{figure*}[htbp]
\addtocounter{figure}{1}
\floatconts
  {fig:gradcam_ad_c1}
  {%
    \addtocounter{figure}{-1}
    \subfigure[AD patient: AD evidence (\textit{int})]{\label{fig:tp_c1_int}%
      \includegraphics[width=0.98\linewidth]{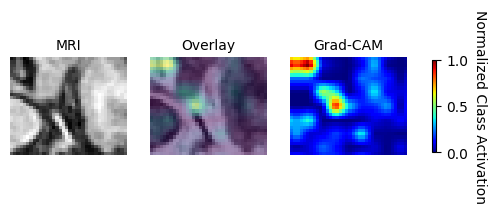}}
    \qquad
    \subfigure[AD patient: AD evidence (\textit{adv})]{\label{fig:tp_c1_adv}%
      \includegraphics[width=0.98\linewidth]{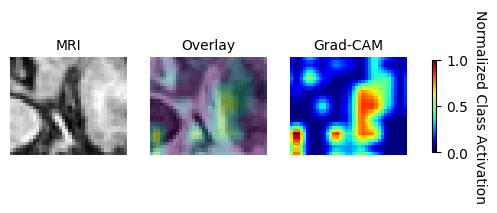}}
  }
{\caption{Representative saliency maps of an \textbf{OOD AD} example. Both the adversarial (\textit{adv}) and the interpretability aware (\textit{int}) model classified this example correctly. The maps highlight the AD evidence. The saliency map associated with the interpretability aware regime highlights focused regions of brain tissue atrophy, such as the gap in the upper left corner. The adversarial saliency map, on the other hand, highlights a large, unfocused area encompassing brain regions that are not clinically meaningful. Improved model robustness against OOD examples therefore seems to translate to more meaningful explanations, too.
  }}
\end{figure*}

% \begin{figure*}[htbp]
% \floatconts
%   {fig:3Dmetrics_5}
%   {%
%     \addtocounter{figure}{-1}
%     \subfigure[TPR]{\label{fig:3D_TPR_5}%
%       \includestandalone[width=0.3\linewidth]{plots/3D/3D_TPR_5}}%
%     \qquad
%     \subfigure[TNR]{\label{fig:3D_TNR_5}%
%       \includestandalone[width=0.3\linewidth]{plots/3D/3D_TNR_5}}
%     \qquad
%     \subfigure[ACC]{\label{fig:3D_ACC_5}%
%       \includestandalone[width=0.3\linewidth]{plots/3D/3D_ACC_5}}
%   }
%   {\caption{Performance metrics shown as a function of $\lambda$. Performance is evaluated on benign ($\epsilon_{test}=0$) and disturbed ($\epsilon_{test}>0$) 2D 3T images, as well as on 1.5T data. All models were trained using $\epsilon = 0.005$, except the `normal' and `combined' model indicated by a circle and triangle/diamond, respectively.
%   }}
% \end{figure*}

% \begin{align}
% L =& \frac{-1}{n}\sum_{i}^{n} y_i \cdot \log \hat{y}_i + (1 - y_i) \cdot \log(1 - \hat{y}_i) \\
% L_{adj} =& L_{adv} + \lambda \tilde{D}_{2, \ell_1}\left(x, x_{adv,\epsilon} \right) \\
% L_{adv} =& \frac{-1}{n}\sum_{i}^{n} y_i \cdot \log \hat{y}_{adv,i} + (1 - y_i) \cdot \log(1 - \hat{y}_{adv,i}) \\
% \tilde{D}_{2, \ell_1}\left(x, x_{adv}\right) =& \frac{1}{2}\left( \|I(x,CN) - I(x_{adv},CN)\|_1 + 
% \|I(x,AD) - I(x_{adv},AD)\|_1 \right)
% \end{align}

\end{document}